\documentclass[twocolumn,aps,pra,notitlepage,floatfix,a4paper,showpacs,preprintnumbers]{revtex4-1}
\usepackage{amssymb, graphicx,subfigure}
\usepackage{enumerate,tensor}
\usepackage{amsmath,bm}
\usepackage{dcolumn}
\usepackage{color}
\usepackage{natbib} 
\usepackage[latin1]{inputenc}
\usepackage{subfigure}
\usepackage{textcomp}
\usepackage{hyperref}
\usepackage{booktabs}
\usepackage{makecell}
\usepackage{psfrag}
\usepackage{epsfig}

\usepackage[per-mode=symbol]{siunitx}   
\usepackage[version=3]{mhchem}

\DeclareSIUnit\intensity{\watt\per\centi\meter\squared}
\DeclareSIUnit\fieldstrength{\volt\per\centi\meter}
\DeclareSIUnit\kfieldstrength{k\volt\per\centi\meter}

\newcommand{\degree}{\ensuremath{^\circ}}%
\newcommand{\eg}{e.\,g.}%
\newcommand{\Estat}{\ensuremath{\textbf{E}_{}{\textup{s}}}}%
\newcommand{\Estatabs}{\ensuremath{\text{E}_{\textup{s}}}}%
\newcommand{\ie}{i.\,e.}%
\newcommand{\Ialign}{\textup{I}_{0}}%
\newcommand{\pstate}[3]{\ensuremath{\left|#1,#2,#3\right>}}%
\newcommand{\ppstate}[3]{\ensuremath{\left|#1,#2,#3\right>_\text{p}}}%
\newcommand{\rppstate}[3]{\ensuremath{_\text{p}\left<#1,#2,#3\right|}}%
\newcommand{\pstateparallel}[3]{\ensuremath{\left|#1,#2,#3\right>^0}}%
\newcommand{\ppstateparallel}[3]{\ensuremath{\left|#1,#2,#3\right>_\text{p}^0}}%

\makeatletter
\def\subsubsection{\@startsection{subsubsection}{3}{10pt}{-1.25ex plus -1ex minus -.1ex}{0ex plus 0ex}{\normalsize\bf}}
\def\paragraph{\@startsection{paragraph}{4}{10pt}{-1.25ex plus -1ex minus -.1ex}{0ex plus 0ex}{\normalsize\textit}}
\renewcommand\@biblabel[1]{#1}
\renewcommand\@makefntext[1]%
{\noindent\makebox[0pt][r]{\@thefnmark\,}#1}
\DeclareRobustCommand\onlinecite{\@onlinecite}
\def\@onlinecite#1{\begingroup\let\@cite\NAT@citenum\citealp{#1}\endgroup}
\def\tagform@#1{\maketag@@@{\ignorespaces#1\unskip\@@italiccorr}}
\let\orgtheequation\theequation
\def\theequation{(\orgtheequation)}
\makeatother

\newcommand{\ket}[1]{|#1\rangle}

\begin{document}
\title{Non-adiabatic effects in long-pulse mixed-field orientation of a linear polar molecule}

\author{Juan J.\ Omiste}
\email{omiste@ugr.es}%
\author{Rosario Gonz\'alez-F\'erez}
\email{rogonzal@ugr.es}%

\affiliation{Instituto Carlos I de F\'{\i}sica Te\'orica y Computacional,
and Departamento de F\'{\i}sica At\'omica, Molecular y Nuclear,
  Universidad de Granada, 18071 Granada, Spain}

\date{\today}
\begin{abstract}
We present a theoretical study of the impact of an electrostatic field combined with non-resonant linearly 
polarized laser pulses on the rotational dynamics of linear molecules. 
Within the rigid rotor approximation, we solve the time-dependent  
Schr\"odinger equation for several field configurations.  
Using the OCS molecule as prototype, the field-dressed dynamics is analyzed in detail 
for experimentally accessible static field strengths and laser pulses. 
Results for directional cosines are presented and compared to the predictions
of the adiabatic theory. 
We demonstrate that  for prototypical field configuration used in current mixed-field orientation experiments,
the molecular field dynamics is, in general, non-adiabatic, being mandatory a time-dependent description
of these systems.  
We investigate several field regimes identifying the sources of non-adiabatic effects, and 
provide the field parameters  under which the adiabatic dynamics would be achieved. 
\end{abstract}
\maketitle

\section{Introduction}
\label{sec:introduction}

The creation of directional states of molecules represents
an important tool to control and tailor
the rotational degree of freedom.  
When a molecule is oriented  the molecular fixed axes are  confined along  laboratory fixed axes and 
its dipole moment is  pointing in a particular direction.
Experimentally, 
the availability of oriented molecules
provides  a wealth of interesting applications in a variety of molecular sciences, 
such as  in chemical reaction dynamics~\cite{brooks:science,brooks:jcp45,loesch:9016,aoiz:chem_phys_lett_289,aquilanti:pccp_7}, 
photoelectron angular distributions~\cite{Bisgaard:Science323:1464,Holmegaard:natphys6,Hansen:PRL106:073001},
 or high-order harmonic generation~\cite{frumker2012,kraus2012}. 
 
 Due to this broad interest, special efforts have been undertaken to create samples of oriented molecules
and techniques based in the application of 
inhomogeneous~\cite{brooks:science,stolte}, and 
homogeneous~\cite{loesch:jcp93,friedrich:nature353,friedrich:jpc95,block:prl68,friedrich:prl69}
electric fields as well as homogeneous magnetic fields~\cite{slenczka:prl72} 
have been used. 
A major breakthrough came with the proposal by
Friedrich and Herschbach~\cite{friedrich:jcp111,friedrich:jpca103}
of enhancing the orientation of polar molecules by exposing
them to  combined weak electrostatic and strong
non-resonant radiative fields.
This theoretical prediction was
done within an adiabatic picture assuming that 
the switching on time of the laser pulse is larger than the molecular rotational period. 
For linear  
molecules, a linearly polarized laser field produces a double-well potential along the
polarization direction. In the pendular limit, this double-well potential contains nearly
degenerate pairs of states with opposite parity forming tunneling doublets. 
If the molecules possess a permanent
electric dipole moment, a strong pseudo-first-order Stark effect is induced by coupling the
tunneling doublets with an additional electrostatic field. 
Due to this coupling, the two levels in a pendular doublet are efficiently oriented but  
with their effective electric dipole moments pointing in opposite directions. 
As a consequence of this oriented and antioriented states pairing,
the orientation is small in a molecular ensemble with the population thermally distributed.
Therefore, the first experimental measures of 
the orientation of a molecular beam were indeed 
reduced to small values~\cite{sakai:prl_90,Buck:IRPC25:583}. 
A significant improvement was gained by using a quantum-state selected molecular beam,
which allowed the creation of unprecedented degree of orientation for complex asymmetric 
tops~\cite{kupper:prl102,ghafur_impulsive_2009,kupper:jcp131}. 
A first theoretical study of the mixed-field orientation experiment of asymmetric top
molecules,  pointed out that a fully adiabatic description of the process does not reproduce
the experimental observations~\cite{omiste:pccp2011}.

Recently, we have experimentally and theoretically investigated the mixed-field orientation  
of the carbonyl sulfide molecule~\cite{nielsen:prl2012}.  
Our analysis has proven that a time-dependent 
description of the mixed-field orientation process is required to explain
the experimental results. 
We have shown how the non-adiabatic coupling of the levels forming the quasi-degenerate doublets 
as the laser intensity is increased, gives rise to the reduction of the orientation 
and, therefore, to the disagreement with the 
predictions of the adiabatic theory~\cite{friedrich:jcp111,friedrich:jpca103}.
Herein, we provide a detailed theoretical analysis
on the dynamics of a linear molecule exposed to an electrostatic
field combined with a non-resonant laser pulse.
In the framework of the rigid rotor approximation, we 
solve the time-dependent Schr\"odinger equation 
using experimental field configurations, \ie, 
a Gaussian laser pulse and a weak electrostatic field that is turned on at constant speed. 
As prototype example, we  consider the carbonyl sulfide molecule (OCS). 
For several rotational states, we  investigate the mixed-field orientation dynamics 
under different field-configurations by varying  either 
the laser peak intensity, 
the duration  of the Gaussian pulse,
the dc field strength or the angle between both fields. 
Hence, we demonstrate that for some field configurations, 
the field-dressed  dynamics is non-adiabatic and
provide a detailed account of the sources of non-adiabaticity and the field regimes at
which they appear. 
For parallel fields,  the  dynamics
is characterized by the population transfer between adiabatic states when the pendular doublets are
formed. Whereas for non-parallel fields, we encounter additional non-adiabatic effects when
the states from the same $J$-manifold, having now the same symmetry,  are
driven apart as the laser intensity is increased on the weak field regime.
For different field configurations, we identify and discuss the experimental conditions needed
to achieve an  adiabatic molecular dynamics.

The paper is organized as follows: In \autoref{sec:hamiltonian_symmetry} we describe the Hamiltonian 
of the system and its symmetries for various field configurations.
The results for the energy, alignment and orientation predicted by the adiabatic theory
are analyzed in \autoref{sec:results_adiabatic}.
In \autoref {sec:results_bneta_0},  we focus on the molecular dynamics when the fields 
are parallel. In particular, we explore how the time-dependent orientation varies as 
the field parameters are modified, and indicate the experimental conditions under
which an adiabatic orientation would be achieved.
A similar study is performed for tilted fields in \autoref{sec:results_beta}, where
we show that the conditions for an adiabatic mixed-field orientation 
are more difficult to fulfill.
In  \autoref{sec:dynamics_top_pulse}, 
we assume that once the pulse is turned on  its peak intensity is kept constant,
and investigate the dynamics in this regime.
In \autoref{sec:ordering},  first the laser pulse is switched on and then an electric field is applied.
In this field configuration, we analyze the orientation of the ground state  and 
provide the field parameters for an adiabatic orientation. 
The conclusions are given in \autoref{sec:conclusions}.

\section{The Hamiltonian of a linear rigid rotor in external fields}
\label{sec:hamiltonian_symmetry}
\begin{figure}[htb]
  \centering
  \includegraphics[width=.25\textwidth]{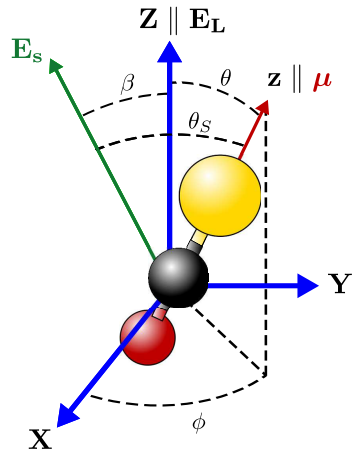}
  \caption{Laboratory fixed coordinate, Euler angles,
schematic field configuration and the OCS molecule.}
  \label{fig:fig_1}
\end{figure}

We consider a polar linear molecule  exposed to an homogeneous
static electric field and a non-resonant linearly polarized laser pulse. 
The field configuration is illustrated in \autoref{fig:fig_1}:
the polarization of the laser field lies along the $Z$-axis of 
the laboratory fixed frame (LFF) $(X,Y,Z)$,
and the dc field is contained in the $XZ$-plane
forming an angle $\beta$ with the $Z$-axis.
The $z$-axis of the molecule fixed frame
$(x,y,z)$ is defined by the permanent dipole moment $\boldsymbol{\mu}$ of the
molecule. These two frames are related by the Euler angles
$\Omega=(\theta,\phi)$, cf \autoref{fig:fig_1}. 
The description of this system is done within the rigid rotor approximation,
assuming that the vibrational and electronic dynamics are not affected by the fields.
Thus, the rigid rotor Hamiltonian reads  
\begin{equation}
  \label{eq:hamiltonian}
  H(t)=H_r+H_{\textup{s}}(t)+H_{\textup{L}}(t),
\end{equation}
where $H_r$ is the field-free Hamiltonian
\begin{equation}
  \label{eq:hr}
  H_r=B\mathbf{J}^2,
\end{equation}
with $\mathbf{J}$ being  the total angular momentum operator and $B$ the
rotational constant. 
The terms $H_{\textup{s}}(t)$ and $H_L(t)$ stands for the
interactions with the static and laser fields, respectively.


The dipole coupling with the static field  reads 
\begin{equation}
  \label{eq:hs}
  H_{\textup{s}}(t)=-\boldsymbol{\mu}\cdot\mathbf{E}_{\textup{s}}(t) 
=-\mu \Estatabs(t)\cos\theta_{\textup{s}}
\end{equation}
with 
$\mathbf{E}_{\textup{s}}(t)=\Estatabs(t)(\sin\beta \hat{X}+\cos\beta \hat{Z})$, 
and $\Estatabs(t)$ being the electrostatic  field strength.
The angle between the dipole moment $\boldsymbol{\mu}$ and this field is 
$\theta_{\textup{s}}$, cf \autoref{fig:fig_1}, and 
$\cos\theta_{\textup{s}}=\cos\beta\cos\theta+\sin\beta\sin\theta\cos\phi$.

We consider a non-resonant laser field linearly polarized along the $Z$-axis of
the LFF,
$\mathbf{E_\textup{L}}(t)=\textup{E}_0g(t)\cos(2\pi\nu t)\hat{Z}$, with 
$\nu$ being its frequency, $\textup{E}_0$ the peak field strength, and $g(t)$ the
pulse envelope. 
Assuming that $\nu^{-1}$ is much shorter than the pulse
duration and the rotational period, 
we average over the rapid oscillations
of the non-resonant field. This causes the coupling of this field with the permanent 
dipole moment to vanish~\cite{dion_pra59,henriksen_cpl312}. 
Thus, the non-resonant laser field molecule interaction 
can be written  as 
\begin{equation}
  \label{eq:hl}
  H_\textup{L}(t)=-\cfrac{\textup{I}(t)}{2c\epsilon_0}\Delta\alpha\cos^2\theta,
\end{equation}
where $\Delta\alpha$ is the polarizability anisotropy, $\textup{I}(t)$ is the intensity of the
laser, $c$ is the speed of light and $\epsilon_0$ is the dielectric
constant. 
Note that we have neglected the term $-\alpha_\perp \textup{I}(t)/2c\epsilon_0$, which represents only a shift in the energy. 

In this work, the field configurations are chosen
based on the mixed-field orientation experiments~\cite{kupper:prl102,kupper:jcp131,nielsen:prl2012}.  
Initially, the molecule is in field-free space, then  
the electrostatic field is switched on, and its
strength is increased linearly with time. 
At $t=-T_0$ the maximum strength is achieved and kept constant afterwards.
This time $T_0$ is chosen long enough to ensure the adiabaticity of this turning-on process.
For the laser pulse, we use a linearly polarized Gaussian pulse with 
a full width half maximum (FWHM) $\tau$ on the nanosecond range.
The intensity  is given by $\textup{I}(t)=\Ialign\exp\left(-t^2/2\sigma^2\right)$, 
with $\Ialign$ being the peak intensity, and  $ \sigma$ is related with the  FWHM $\tau=2\sqrt{2\ln 2} \sigma $.  
Numerically, the non-resonant laser field is turned on in such a way that   
the interaction due to this field is much weaker 
than coupling with  the dc field.

\begin{table}[t]  
\caption{Action of the symmetry operations on the Euler angles.}
\label{table:symmetry}  
\begin{ruledtabular}
  \begin{tabular}{lll}
    &Transformations&\\
    \hline
    Operation\;&$\phi$ &$\theta$\\
    \hline
    ${\cal E}$\,&$\phi\rightarrow \phi$\,&$\theta\rightarrow \theta$\\
    $\sigma_{XZ}$\,&$\phi\rightarrow2\pi-\phi$\,&$\theta\rightarrow\theta$\\
    ${\cal C}_X(\pi)$\,&$\phi\rightarrow2\pi-\phi$\,&$\theta\rightarrow\pi-\theta$\\
    ${\cal C}^\alpha_{\perp Z}(\pi)$\,&$\phi\rightarrow2\alpha-\phi$\,&$\theta\rightarrow\pi-\theta$\\
    ${\cal C}_Z(\delta)$\,&$\phi\rightarrow\phi+\delta$\,&$\theta\rightarrow \theta$\\
  \end{tabular}
\end{ruledtabular}
\end{table}

The eigenstates of the field-free Hamiltonian \eqref{eq:hr}
are the spherical harmonics $Y_{JM}(\Omega)$, 
with $J$ and $M$ being  the rotational and magnetic quantum numbers, respectively.
Note that $M$ is the projection of the total angular momentum $\mathbf{J}$
on the   LFF $Z$-axis.
The field-free Hamiltonian \eqref{eq:hr}
belongs to the SO(3) group because the operator
$\mathbf{J}^2$ remains unaltered under any
rotation. 
In the presence of the external fields, the symmetries of the rotational Hamiltonian
\eqref{eq:hamiltonian} are significantly reduced.
The operations describing  the symmetries
of  the field-dressed Hamiltonian are collected in \autoref{table:symmetry}. 
For a static field, 
the symmetry  group is reduced to arbitrary rotations around the field axis and the
identity $\{{\cal E}, {\cal C}_{\Estat}(\delta)\}$. If only a linearly polarized laser field is
applied, the Hamiltonian is invariant under arbitrary rotations around $Z$-axis and  
two-fold rotations around any axis perpendicular to the $Z$-axis,
and the symmetry group is compound by $\{{\cal E}, {\cal C}_{Z}(\delta), {\cal C}_{\perp Z}(\pi)\}$.  
For parallel fields, the elements of the symmetry  group are
the identity ${\cal E}$, arbitrary rotations around the fields ${\cal C}_{Z}(\delta)$ and
the reflection in any plane containing the fields. 
For a given $|M|$-value, the parity under the reflection on one of these planes defines  two irreducible
representations. Since the selection of this plane is not unique, 
then the states with  $M\ne 0$ are doubly degenerated.
Note that $M$ remains as a good quantum number for these three field configurations 
with $\beta=0\degree$. 
For non-perpendicular and non-collinear fields, i.e., $\beta\ne 0^{\circ}$ and $\beta\ne 90^{\circ}$, the 
Hamiltonian is invariant under the 
identity ${\cal E}$ and the reflection on the  $XZ$-plane 
containing the fields $\sigma_{XZ}$. This group $\{{\cal E}, \sigma_{XZ} \}$   
has only two irreducible
representations characterized by the parity with respect to this reflection $\sigma_{XZ}$, and
the functions belonging to the even and odd representations are 
$ \Psi_{JM}^{e/o}(\Omega)=\left(Y_{JM}(\Omega)+(-1)^{\kappa}Y_{J-M}(\Omega)\right)/\sqrt{2}$,
with $\kappa$ given in \autoref{tab:irrep_xzperp}. 
If the fields are perpendicular, $\beta=90^{\circ}$, 
the two-fold rotation around the static
field ${\cal C}_X(\pi)$ is also a symmetry operation. Thus, the symmetry group
is formed by 
$\{{\cal E}, \sigma_{XZ}, {\cal C}_X(\pi)\}$  has 
four irreducible representations labeled by the parity with respect to
the transformations $\sigma_{XZ}$ and ${\cal C}_X(\pi)$.
The properly symmetrized functions for these four irreducible representations are
$ \Psi_{JM}^{e/o,e/o}(\Omega)=\left(Y_{JM}(\Omega)+(-1)^{\epsilon}Y_{J-M}(\Omega)\right)/\sqrt{2}$,
the possible values of $\epsilon$ are collected in \autoref{tab:irrep_xzperp}. 
\begin{table}[htp]
  \centering
  \caption{For the 
$\beta=90\degree$ and  $0\degree<\beta<90\degree$ field configurations,
  values of the parameters  $\kappa$ and $\epsilon$ for the 
wave functions with the correct symmetry for each irreducible
representation.} 
  \label{tab:irrep_xzperp}
\begin{ruledtabular}
  \begin{tabular}{cccccc}
 \multicolumn{2}{c}{$\beta\ne90\degree$}&&{$\beta=90\degree$}\\
\cline{1-2} \cline{3-6} 
      $\kappa$ &$\sigma_{XZ}$&  $\epsilon$ &$J+M$& $\sigma_{XZ}$ & ${\cal C}_X$ 
 \\
\hline
  $M$ &even& $M$& even & even & even\\
  $M+1$ &odd&$M$& odd & even & odd \\
 &&${M+1}$ & odd & odd & even  \\
&&${M+1}$ & even & odd & even\\
  \end{tabular}
\end{ruledtabular}
\end{table}

The time-dependent Schr\"odinger equation 
associated to the Hamiltonian \eqref{eq:hamiltonian}
is solved by means of 
a second-order split-operator technique~\cite{feit:jcp82}, combined with 
the discrete-variable and finite-basis representation
methods for the angular coordinates~\cite{bacic:arpc89,corey:jcp92,offer:10416,sanchezmoreno:PRA.2007}. 
For reasons of addressability, we will label the 
time-dependent states as \pstateparallel{J}{M}{l} and \pstate{J}{M}{l}
for $\beta=0\degree$ and $0\degree<\beta<90\degree$, 
respectively, with $l=e$ and $o$ indicating even or odd parity
with respect to the $XZ$-plane.
The labels $J$ and $M$ refer to field-free quantum number to which they are adiabatically
connected. 
Note that the labeling of the states depends on the way the fields are turned on~\cite{hartelt_jcp128}. 
The time-dependent wave function depends on the time $t$, 
the peak intensity $\Ialign$, the FWHM $\tau$, the electrostatic field strength $\Estatabs$,
and the  angle $\beta$. 
For the  sake of simplicity, we have not made 
explicit these dependences, but the
field configuration is clearly indicated through the text. 

To get a better physical insight on the field-dressed dynamics, 
the time-dependent results will be compared to those from
the adiabatic theory. 
For this system, 
we take the adiabatic limit 
by using a constant electrostatic field $\Estatabs$
and constant laser intensity $\textup{I}$ in the Hamiltonian \eqref{eq:hamiltonian}.
The time-independent Schr\"odinger equation associated
to this Hamiltonian is solved by expanding the wave function in a basis that respects
the symmetries.
The adiabatic states are labeled as
\ppstateparallel{J}{M}{l} and \ppstate{J}{M}{l}
for $\beta=0\degree$ and $0\degree<\beta<90\degree$, 
respectively, and 
we have  not made explicit their
dependence on the field parameters. 

The field-dressed eigenfunctions of this time-independent Hamiltonian 
form a basis, which is used to  analyze
the time-dependent wave function
$| \gamma\rangle$  
by means of the following expansion
\begin{equation}
 \label{eq:projection_psit}
 |\gamma\rangle=\sum_{i=1}^NC_i(t)|\gamma_i\rangle_\textup{p}
\end{equation}
with $  C_i(t)=\langle \gamma|\gamma_i\rangle_\textup{p}$,  
$\gamma$ and $\gamma_i$ including all the 
labels identifying these levels.
For computational reasons, we have only considered the lowest-lying  $N$ adiabatic levels, 
and always ensured that the contributions of highly excited states are 
negligible.
Let us remark that for each time $t$, the expansion of the wave function is performed 
in a different adiabatic basis obtained by solving
the time-independent  Schr\"odinger equation using the
 static field strength and laser intensity at time $t$, i.e.,  $\Estatabs(t)$ and $\textup{I}(t)$.

\section{Results in the adiabatic limit}
\label{sec:results_adiabatic}

In this work, we use 
the OCS molecule (see fig. \ref{fig:fig_1}) as benchmark
 to illustrate our results. 
The rotational constant of OCS is $B=0.209$ cm$^{-1}$, 
the permanent dipole moment $\mu=0.709$~D and the polarizability anisotropy
$\Delta\alpha=4.04$ \AA$^3$.

\begin{figure}[htb]
  \centering
  \includegraphics[width=.4\textwidth,angle=0]{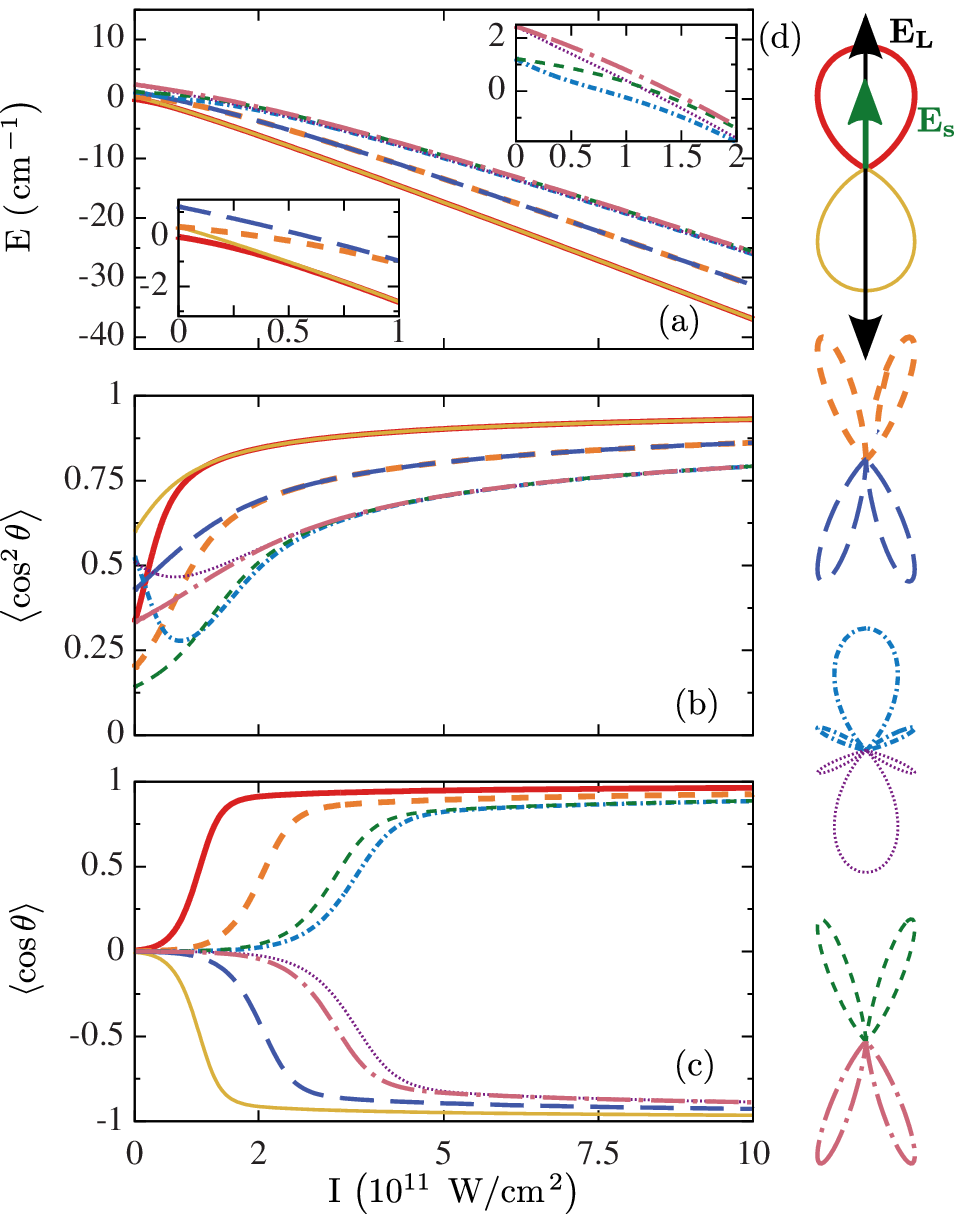}
   \caption{(Color online) Adiabatic results for the (a) energy,
expectation values (b) $\langle\cos^2\theta\rangle$ and
(c) $\langle\cos\theta\rangle$ as a function $\textup{I}$
of the adiabatic states   
\ppstateparallel{0}{0}{e} (red thick solid),  
\ppstateparallel{1}{0}{e} (gold thin solid),   
\ppstateparallel{1}{1}{e} (orange thick short-dashed), 
\ppstateparallel{2}{1}{e} (dark blue long-dashed),   
\ppstateparallel{2}{0}{e} (blue dot-short-dashed),  
\ppstateparallel{3}{0}{e} (purple dotted), 
\ppstateparallel{2}{2}{e} (green thin short-dashed),  
and
\ppstateparallel{3}{2}{e} (pink dot-long-dashed).
The insets show the relevant energy and intensity ranges where the formation
of the near-degenerate doublets occurs. 
(d) Polar plots  of the square of their wave functions at $\textup{I}=\SI{e12}{\intensity}$. 
$\Estatabs=\SI{300}{\fieldstrength}$  and $\beta=0\degree$ for all data.}
  \label{fig:fig_2}
\end{figure}
We start by analyzing the 
adiabatic limit. 
We restrict this study to the following 
eight states: 
\ppstateparallel{0}{0}{e},  
\ppstateparallel{1}{0}{e}, 
\ppstateparallel{1}{1}{e}, 
\ppstateparallel{2}{1}{e}, 
\ppstateparallel{2}{0}{e}, 
\ppstateparallel{3}{0}{e},
\ppstateparallel{2}{2}{e}, 
and
\ppstateparallel{3}{2}{e}. 
For $\beta=0\degree$, 
they
adiabatically correspond to the 
states forming the four first doublets. 
Note that they well represent the main physical features observed in
the overall molecular dynamics, and similar behavior and properties are,
therefore, obtained for states in  other irreducible representations. 

For $\Estatabs=\SI{300}{\fieldstrength}$ and $\beta=0\degree$, 
the energies and the expectation values  $\langle\cos^2\theta\rangle$ and $\langle\cos\theta\rangle$
of these levels are plotted versus the laser intensity in
\autoref{fig:fig_2}~(a), (b) and (c), respectively. 
The weak static field breaks the field-free degeneracy in the magnetic quantum number and, as
the laser intensity is increased, these states become  high field seekers. 
In the strong laser field regime, once the pendular regime is reached, pairs of quasi-degenerate
states with the same symmetry are formed. The insets in this panel show how these doublets appear.   
The gap in energy in a doublet goes as 
$\Delta E\approx2\left|\mu \Estatabs \, {}_\textup{p}\langle i|\cos\theta|i\rangle_\textup{p} \right|$, where
$\mu \;{}_\textup{p}\langle i|\cos\theta|i\rangle_\textup{p}$ is the effective dipole moment of the state 
$|i\rangle_\textup{p}$ in the doublet, which is of opposite sign  for $|j\rangle_\textup{p}$.
Within a doublet, the two levels are characterized by the same hybridization of the angular motion 
$\langle \mathbf{J}^2\rangle$ and alignment $\langle\cos^2\theta\rangle$, see  \autoref{fig:fig_2}~(b).
In contrast, they possess  opposite orientation $\langle\cos\theta\rangle$, one being oriented and the other
antioriented, cf. \autoref{fig:fig_2}~(c).
This opposite orientation  is also illustrated in \autoref{fig:fig_2}~(d)  by 
the polar plots of the square of their wavefunctions for $\textup{I}=\SI{e12}{\intensity}$.  
The larger is the field-free rotational quantum number of the levels, \ie,
their field-free energy, the stronger is the laser intensity needed to achieve a
significant orientation.
Indeed, the states in the third and fourth doublets
are not aligned for  $\textup{I}\lesssim\SI{2e11}{\intensity}$ 
and, therefore, not oriented. 
Once the pendular regime is achieved,
the orientation of these states $|\langle\cos\theta\rangle|$ 
approaches to $1$ as $\textup{I}$ is enhanced. 
If the laser field is sufficiently strong, this adiabatic orientation is 
independent of the dc field strength, and of the angle between both fields.

\section{Results for parallel fields}
\label{sec:results_bneta_0}

\begin{figure}[htb]
  \centering
  \includegraphics[width=.45\textwidth,angle=0]{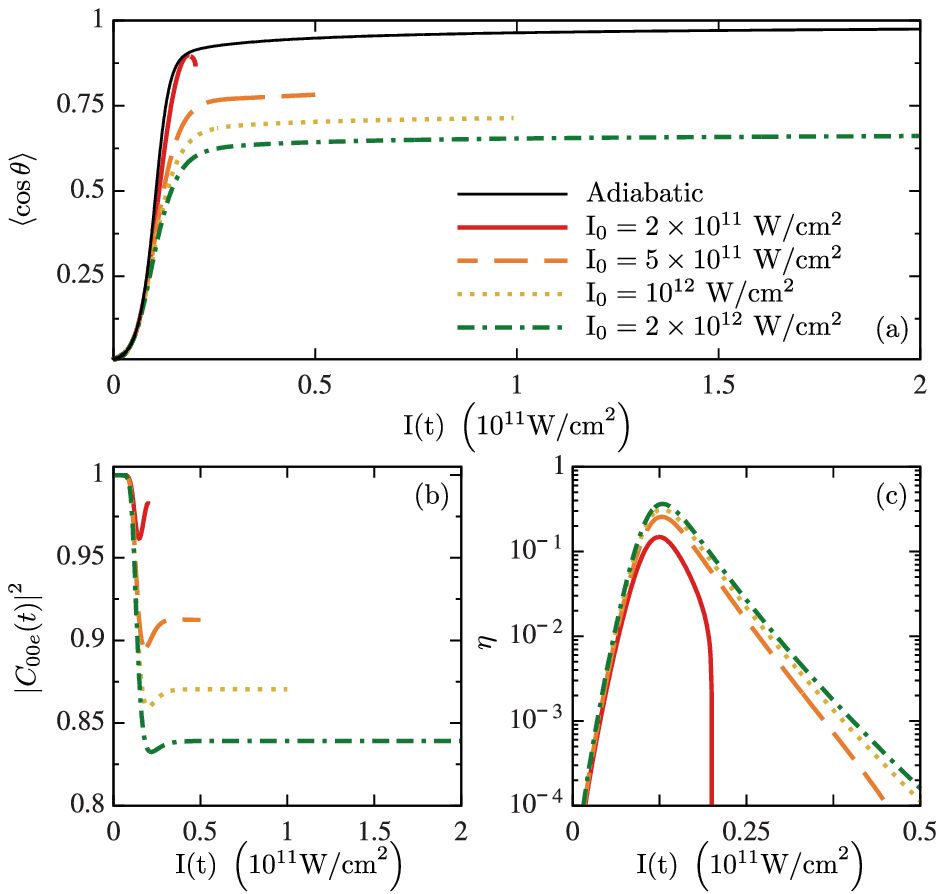}
  \caption{(Color online)
(a) For the ground state, 
time evolution of the expectation value $\langle\cos\theta\rangle$  as a function of $\textup{I}(t)$
for Gaussian pulses of $\tau=10$~ns and peak intensities
$\Ialign=\SI{2e11}{\intensity}$ (red thick  solid),  
$\SI{5e11}{\intensity}$ (orange dashed), 
$\SI{e12}{\intensity}$ (gold dotted)
and
$\SI{2e12}{\intensity}$ (green dot-dashed).
The adiabatic results for  $\langle\cos\theta\rangle$ (thin solid) are also included.
(b) The squares of the projections of the time-dependent wave functions
onto the adiabatic pendular state \ppstateparallel{0}{0}{e}. 
(c) Adiabatic criteria $\eta$ as a function of $\textup{I}(t)$. 
The field configuration is
$\Estatabs=\SI{300}{\fieldstrength}$ and $\beta=0\degree$.} 
  \label{fig:fig_3}
\end{figure}

In this section,  we investigate the rotational dynamics 
in a parallel configuration: a dc-field of $\SI{300}{\fieldstrength}$ 
and a Gaussian pulse with FWHM $\tau=10$~ns and several peak intensities. 
For the ground state \pstateparallel{0}{0}{e}, 
the expectation value $\langle\cos\theta\rangle$
is presented in \autoref{fig:fig_3}~(a) 
as a function of $\textup{I}(t)$ up until the peak intensity $\Ialign$ is reached. For comparison,
the adiabatic results are also shown.  

Since the FWHM is 125.31 times larger than the rotational period, one would expect that the 
rotational dynamics follows the adiabatic limit. However this is not the case, and there are
significant discrepancies between the  time-dependent 
and adiabatic results. 
In contrast  to what is predicted by the adiabatic theory, the final orientation
decreases as the peak intensity of the laser pulse  is increased. 
For $\Ialign=\SI{2e11}{\intensity}$,  $\langle\cos\theta\rangle$ initially  resembles the adiabatic
behavior,  but it achieves a maximum value $\langle\cos\theta\rangle=0.899$ for
$\textup{I}(t)=\SI{1.86e11}{\intensity}$. 
For $\Ialign=\SI{5e11}{\intensity}$, $\SI{e12}{\intensity}$, and $\SI{2e12}{\intensity}$,
the orientation shows a qualitatively similar but
quantitatively different behavior: 
in the weak laser field regime, $\langle\cos\theta\rangle$ monotonically increases  following the adiabatic limit, 
but for $\textup{I}(t)\gtrsim\SI{2e11}{\intensity}$  it reaches a 
plateau behavior being the orientation smaller than the
adiabatic value. For instance,
$\langle\cos\theta\rangle=0.661$ for $\textup{I}(0)=\SI{2e12}{\intensity}$
whereas the adiabatic value is $\langle\cos\theta\rangle=0.975$. 

A first physical insight into  the non-adiabatic dynamics could
be gained  by analyzing the characteristic times of the  molecule. 
When the states in a pendular doublet are quasi-degenerated, 
the energy gap between them,  $\Delta E\sim 2\mu \Estatabs$, 
defines a time scale of this system~\cite{nielsen:prl2012}. 
Note that 
we have assumed $|{}_\textup{p}\langle i|\cos\theta|i\rangle_\textup{p}|=|{}_\textup{p}\langle j|\cos\theta|j\rangle_\textup{p}|\approx 1$,
which holds in the strong laser field regime. 
For $\Estatabs=\SI{300}{\fieldstrength}$ and $\beta=0\degree$,
the energy separation
within the first doublet formed by \ppstateparallel{0}{0}{e}  and \ppstateparallel{1}{0}{e}  
is $\Delta E=6.97\times 10^{-4}$~cm$^{-1}$  giving a time scale
of $761.21$~ps, which is larger than the rotational period 
$79.8$~ps.
Thus, only long enough pulses compared to this pendular time would
ensure an adiabatic orientation of the molecule.

Since the static field strength is so weak,  its impact on the rotational dynamics is very small, and 
before the pulse 
the levels could be considered as field-free rotor states.
As the laser intensity increases, the states are hybridized by the combined action of the both fields,
and  the doublets of nearly-degenerate states are formed in the
strong laser field regime, as it is shown in  \autoref{fig:fig_2}~(a).
When the energy splitting of this pendular doublet 
approaches the coupling of the two sublevels 
due to  the  pseudo-first-order Stark interaction, 
these states can mix because they
have the same symmetry for $\beta\ne90\degree$.
As a consequence, there is 
a population transfer between the oriented and anti-oriented states, which results in a decrease
of the final orientation compared to the adiabatic limit.  
For this field configuration, the dynamics can be analyzed by means of the adiabatic states
forming this pendular doublet, because 
their couplings to states in  neighboring  doublets is much smaller
than the energy difference between them. 
Note that these adiabatic states are the eigenstates of the Hamiltonian at fixed time $t$. 


Under a time-dependent interaction, \ie, in our case 
the interaction with the laser field $H_\textup{L}(t)$ 
\eqref{eq:hl}, 
the dynamics could be considered as adiabatic if  and only if 
the following condition~\cite{ballentine:quantum_mechanics} 
\begin{equation}
  \label{eq:adiabatic_criteria}
  \eta=\cfrac{\hbar\left| 
\tensor[_{\textup{p}}]{\left\langle i \left|\cfrac{\partial H_\textup{L}(t)}{\partial t}\right| j\right\rangle}{_{\textup{p}}}
\right|}{\left|E_i-E_j\right|^2}\ll 1
\end{equation}
is fulfilled, with $\ket{i}_\textup{p}$ and $\ket{j}_\textup{p}$ being  the eigenstates of the Hamiltonian in the adiabatic limit, and 
$E_i$ and $E_j$ their energies.
According to this criterion,  the probability for mixing,
corresponding to a transfer of population
from one state of the doublet to the other, is
determined by the rate of change of the laser field interaction
 and the energy separation between the states.
Thus, as the laser intensity is increased
the population transfer between the two states in a doublet takes place because
the criterion \eqref{eq:adiabatic_criteria} is not satisfied.
To illustrate this phenomenon, we show the contribution of the adiabatic ground state 
$|C_{00e}(t)|^2$ to the 
time-dependent wave function of \pstateparallel{0}{0}{e}, \autoref{fig:fig_3}~(b), and the
adiabatic parameter 
$\eta$ when $\eta\ge 10^{-4}$, \autoref{fig:fig_3}~(c).
Note that $|C_{00e}(t)|^2+|C_{10e}(t)|^2=1$. 
In these four cases, the dynamics is  initially adiabatic, \ie,  $|C_{00e}(t)|^2$ remains equal to $1$ and $\eta\ll 1$.
As $\textup{I}(t)$ is increased, the energy splitting of the doublet decreases and, moreover, it becomes comparable or even larger 
than the rate of turning-on the pulse; thus, $\eta$ is close to $1$, and the population transfer
takes place.
This region where $\eta$ is not negligible
corresponds to the formation of the quasi-degenerate
doublet.
Once the doublet is formed,   $\Delta E=|E_i-E_j|$ reaches a small value and slowly decreases as
$\textup{I}(t)$ is enhanced; 
but the two states are oriented in opposite directions and their wave functions do not overlap.
Therefore, the coupling due to the alignment laser is much smaller than
$\Delta E$,
 $\eta\ll1$  and the population transfer
does not take place any longer, i.e.,  $|C_{00e}(t)|^2$ remains constant as
$\textup{I}(t)$ is  enhanced.
The larger is this population transfer, the smaller 
is the orientation compared to the adiabatic prediction. 
Since these adiabatic states contributing to the dynamics are quasi-degenerated
and have very close values of the alignment and hybridization of the angular motion, 
the lack of adiabaticity is not reflected on 
the time evolution of the energy, $\langle\cos^2\theta\rangle$ or $\langle\mathbf{J}^2\rangle$.

For this field configuration,
the molecular dynamics of excited states present
analogous features  for $\langle\cos\theta\rangle$,
$\langle\cos^2\theta\rangle$ and $\langle\mathbf{J}^2\rangle$ 
as those encountered here for  the ground state.

The adiabaticity of the field-dressed dynamics is determined by the
rate of change of the laser field  
interaction compared to the largest time scale of the system.
In the pendular regime, 
the energy splitting in a doublet goes as $\Delta E\sim 2\mu  \Estatabs$; 
then, 
the population transfer decreases if $\Estatabs$ 
is increased.   
On the other hand, by  increasing the FWHM augments the time scale on which
the pendular doublets are formed, and facilitates the adaptation of the molecule 
to this field. 
That is, the mixed-field orientation will be more adiabatic 
when either  longer pulses or stronger static electric fields are used.
Let us remark that with the expression {\it being the dynamics more adiabatic} we mean that
for a certain state,  
the weight of its corresponding adiabatic state in the time-dependent wave function
is closer to one during the dynamics.

 \subsection{Influence of the peak intensity $\Ialign$}
\label{sec:intensity}

\begin{figure}[h]
  \centering
  \includegraphics[width=.47\textwidth]{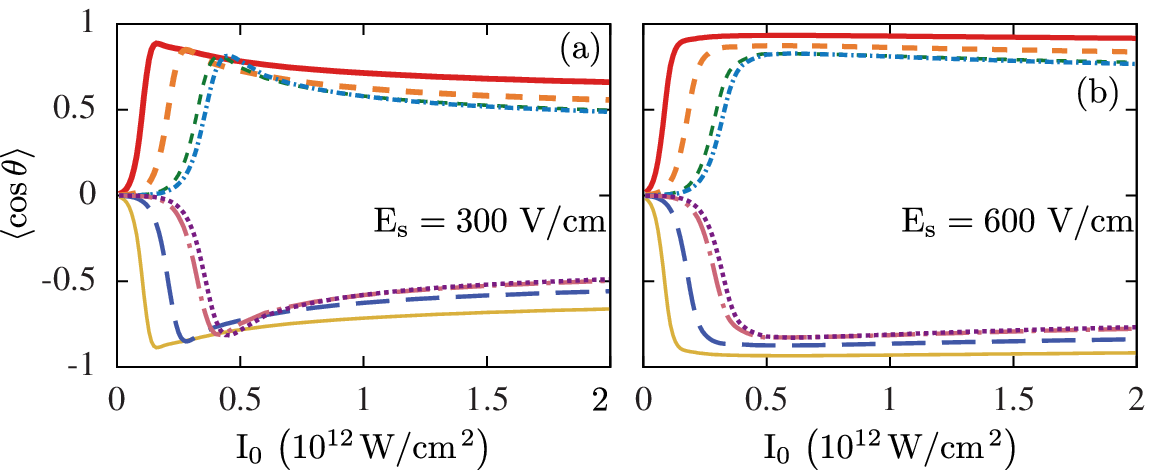}
  \caption{(Color online) 
Expectation value $\langle\cos\theta\rangle$ at $t=0$  as a function of 
the peak intensity $\Ialign$ for  the states
\pstateparallel{0}{0}{e} (red thick solid), 
\pstateparallel{1}{0}{e} (gold thin solid)
\pstateparallel{1}{1}{e} (orange thick short-dashed),
\pstateparallel{2}{1}{e} (dark blue long-dashed)
\pstateparallel{2}{0}{e} (blue dot-short-dashed), 
\pstateparallel{3}{0}{e} (purple dotted),
\pstateparallel{2}{2}{e} (green thin short-dashed) and
\pstateparallel{3}{2}{e} (pink dot-long-dashed), 
for $\beta=0\degree$ and 
(a) $\Estatabs=\SI{300}{\fieldstrength}$ and (b)
$\SI{600}{\fieldstrength}$.
The FWHM of the laser pulses is $10$~ns.}
  \label{fig:fig_4}
\end{figure}

Here, we investigate the orientation at the maximum of the laser
pulse, as it is done in most of the experiments
\cite{sakai:prl_90,friedrich:jmodopt50,kupper:jcp131}. 
The rate of change of the laser field  and the adiabatic parameter
\eqref{eq:adiabatic_criteria} depend linearly on $\Ialign$. 
Then, for a Gaussian pulse with fixed FWHM, the dynamics will 
be more diabatic if $\Ialign$ is increased.  
In this section, we consider the  states 
\pstateparallel{0}{0}{e},  
\pstateparallel{1}{0}{e}, 
\pstateparallel{1}{1}{e}, 
\pstateparallel{2}{1}{e}, 
\pstateparallel{2}{0}{e}, 
\pstateparallel{3}{0}{e}, 
\pstateparallel{2}{2}{e}, 
and 
\pstateparallel{3}{2}{e}.
Their  orientation at $t=0$, \ie,  $\langle\cos\theta\rangle$
for $\textup{I}(0)=\Ialign$, is plotted as a function of $\Ialign$ in  \autoref{fig:fig_4}~(a) and
(b) for $\Estatabs=\SI{300}{\fieldstrength}$ and $\SI{600}{\fieldstrength}$, respectively.
The fields are parallel, and 
the FWHM of these pulses is $10$~ns. 

For $\Estatabs=\SI{300}{\fieldstrength}$, the orientation  of the low-lying level in
a doublet increases
as $\Ialign$ is enhanced, reaching a maximum  and smoothly decreasing thereafter. 
This is counterintuitive to what is expected in the adiabatic limit; namely, 
a larger orientation when the laser intensity is increased. 
The maximum in the orientation is achieved with an alignment pulse 
that already gives rise to a non-adiabatic dynamics. However, 
due to the coupling between
the populated adiabatic states in the  pendular pair
${}_\textup{p}\langle i|\cos\theta|j\rangle_\textup{p}$, 
the orientation is enhanced compared to what happens at the adiabatic limit.  
By further increasing $\Ialign$, the population transferred between the
two states is enhanced, but now the coupling between them 
is very small or even zero due to their opposite orientation. As a consequence, 
the final orientation decreases as  
$\Ialign$ is increased. 
For a certain pendular doublet, the upper state is antioriented, and $\langle\cos\theta\rangle$
shows the opposite  behavior as a function of $\Ialign$.
Regarding the third and fourth doublets,
the states are not oriented nor
aligned for ac pulses with $\Ialign\lesssim \SI{2.4e11}{\intensity}$. 
Compared to low-lying states, their orientation is smaller and the maximum of $\langle\cos\theta\rangle$
appears at larger peak intensities.

By increasing the static field strength  to $\SI{600}{\fieldstrength}$
the energy gap of the pendular pair is also increased,  
being the characteristic time scale of the system reduced. 
Thus, for the same laser pulse, the dynamics is more adiabatic, \ie,
less population is transferred, and 
the final orientation is increased, see \autoref{fig:fig_4}~(b). 
The orientation (antiorientation) of the
pendular states also achieves a maximum (minimum), but it is so
shallow that it is hardly appreciated on the scale of this figure.

\begin{figure}[h]
  \centering
  \includegraphics[angle=0,width=.48\textwidth]{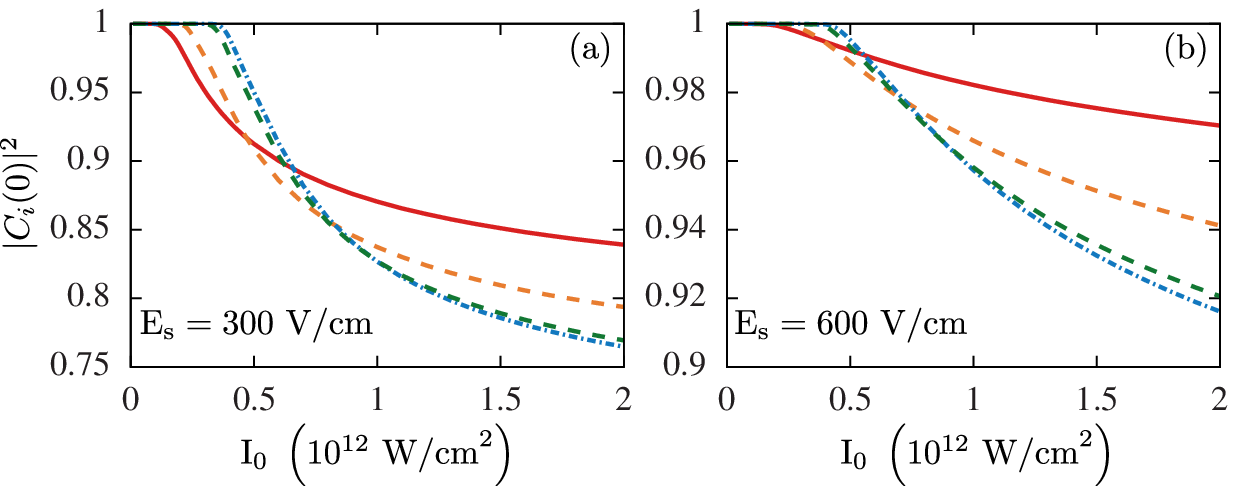}
  \caption{(Color online) 
Projections of the time-dependent wave functions onto the 
corresponding adiabatic states as a function of 
the peak intensity $\Ialign$ for  the states
\pstateparallel{0}{0}{e} (red thick solid), 
\pstateparallel{1}{1}{e} (orange thick short-dashed),
\pstateparallel{2}{0}{e} (blue dot-short-dashed)
and 
\pstateparallel{2}{2}{e} (green thin short-dashed), 
for (a) $\Estatabs=\SI{300}{\fieldstrength}$ and 
(b) $\SI{600}{\fieldstrength}$.
We use $10$~ns laser pulses and $\beta=0\degree$.}
  \label{fig:fig_5}
\end{figure}
To illustrate the field-dressed dynamics, 
we plot in \autoref{fig:fig_5} the weights of the adiabatic states associated to
the oriented levels in these pendular doublets.
For the corresponding antioriented levels, the contributions of its associated adiabatic state are identical to
the one presented here,
\eg, 
for the ground state,
we present the contribution of the adiabatic ground state $|C_{00e}(0)|^2$,
which is identical to the weight $|C_{10e}(0)|^2$ for \pstate{1}{0}{e}. 
In an adiabatic molecular dynamics, these coefficients
are  equal to one. 
Note that in the considered regime 
only the two adiabatic states of the pendular doublet contribute to the dynamics.
For all these levels,  $|C_i(0)|^2$ decreases, \ie, the dynamics is less adiabatic, as
$\Ialign$ is enhanced. By increasing $\Estatabs$,  $\Delta E$ is increased; 
thus, under the same Gaussian pulse
the population transfer is reduced, \ie,  $|C_i(0)|^2$ is closer to one, and
the range of peak intensities that could be considered as adiabatic is increased.

\subsection{Influence of the FWHM of the laser pulse}
\label{sec:width}
\begin{figure}[h]
 \centering
  \includegraphics[width=.5\textwidth]{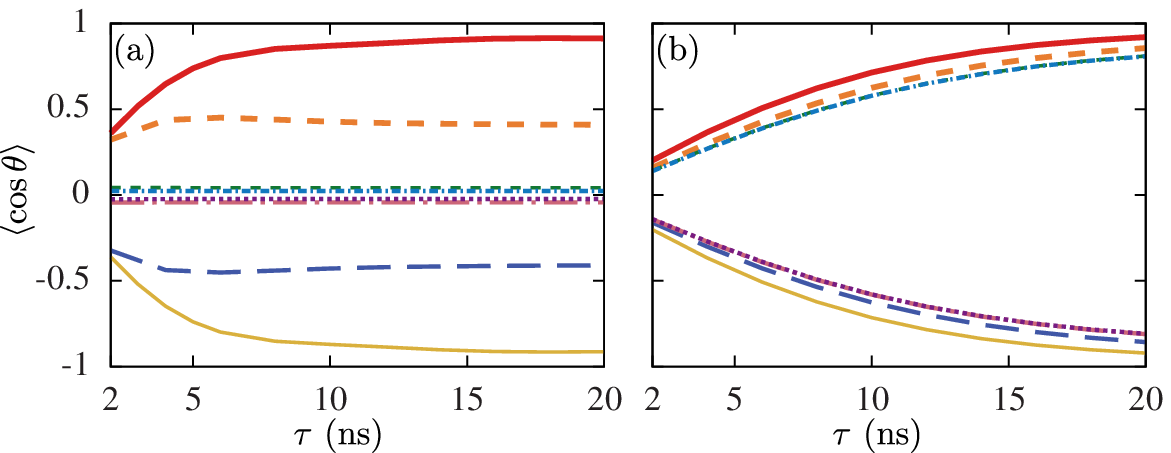}
\caption{(Color online)  
Expectation value  $\langle\cos\theta\rangle$ 
at $t=0$ as a function of 
$\tau$ for 
(a)  $\Ialign=\SI{2e11}{\intensity}$
and 
(b) $\Ialign=\SI{e12}{\intensity}$.  
The fields 
are parallel  and  $\Estatabs=\SI{300}{\fieldstrength}$.
The states and their labels are the same as in
  \autoref{fig:fig_4}.}
  \label{fig:fig_6}
\end{figure}

The duration of the Gaussian pulse plays an important role in the 
molecular dynamics. 
It has been shown that an alignment pulse of $10$~ns is not enough
to achieve an adiabatic mixed-field orientation
for molecules such as OCS, benzonitrile and 
iodobenzene~\cite{poulsen:phys_rev_a_73,friedrich:jpca103,hartelt_jcp128,nielsen:prl2012}.  
It has been pointed out the need of increasing the rising time of
the laser pulses to achieve the highest possible orientation~\cite{Sugawara2008,Muramatsu2009,nielsen:prl2012}.   
By increasing the FWHM, 
the time needed to form the pendular doublets is also increased.
For a given field configuration, 
at the point where the pulse reaches a certain intensity 
the adiabatic parameter \eqref{eq:adiabatic_criteria} 
is reduced if $\tau$ is increased.
Hence, the molecular dynamics becomes more adiabatic and, therefore, the
population transfer to other pendular states is reduced. 
Here, we investigate how the directional properties of OCS depends on the
laser-pulse FWHM.
For the same set of states as in the previous section,
\autoref{fig:fig_6} 
shows the orientation  at $t=0$  as a
function of $\tau$. The fields are parallel, and we consider 
the peak intensities  
$\Ialign=\SI{2e11}{\intensity}$ and $\SI{e12}{\intensity}$, and a dc field of 
$\Estatabs=\SI{300}{\fieldstrength}$.

The degree of orientation of the two states in a given pendular pair shows the 
same behavior as a function of $\tau$, but with their dipole moment  pointing in opposite directions. 
In the first two doublets and $\Ialign=\SI{2e11}{\intensity}$, $|\langle\cos\theta\rangle|$ increases
with $\tau$ till it reaches a plateau-like behavior.  
The second pair satisfies that $|\langle\cos\theta\rangle|\approx0.428$  for
$\tau\gtrsim10$~ns. 
Since the states in the third and fourth doublets have not achieved the pendular regime  
for $\Ialign=\SI{2e11}{\intensity}$, increasing the pulse duration does not have any impact on 
their orientation, and $\langle\cos\theta\rangle$ keeps a constant value close to zero as $\tau$ is
increased. 
For $\Ialign=\SI{e12}{\intensity}$,
the degree of orientation of all the states increases
and approaches the adiabatic limit as $\tau$ is enhanced. 
For the ground state and $\tau=20$~ns, we obtain $\langle\cos\theta\rangle=0.913$,
which is very close to the adiabatic value $\langle\cos\theta\rangle=0.964$.

These results show that for parallel fields, 
the mixed-field orientation dynamics
of any state could be adiabatic 
if a sufficiently long pulse and sufficiently strong fields are used. 
For a $50$~ns Gaussian pulse with $\Ialign=\SI{e12}{\intensity}$,
and $\Estatabs=\SI{300}{\fieldstrength}$,   the dynamics can be considered as 
adiabatic for all these states, with $|C_i(0)|^2\gtrsim 0.999$.

\subsection{Influence of the electrostatic field strength}
\label{sec:static}

\begin{figure}[h]
  \centering
  \includegraphics[width=.5\textwidth]{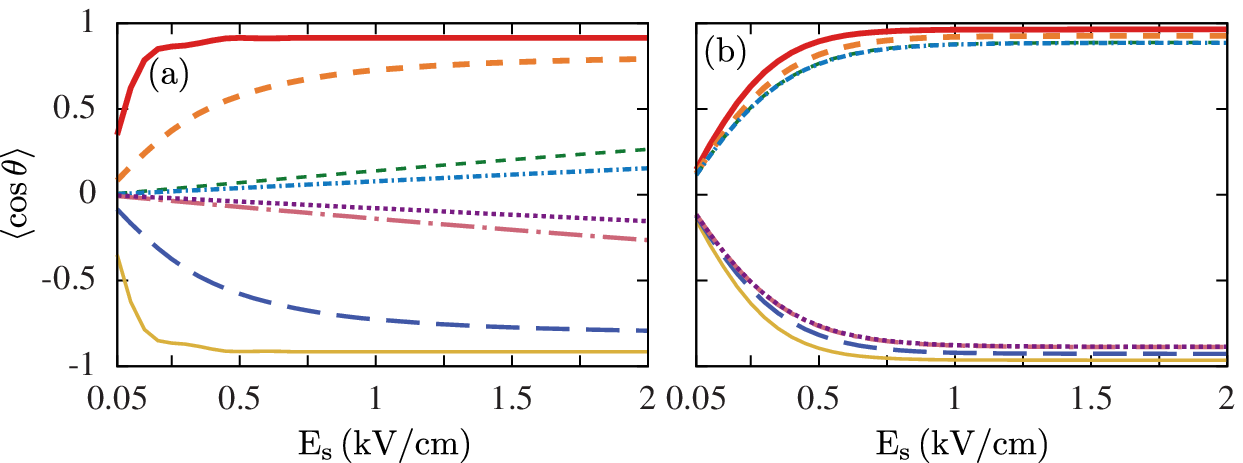}
  \caption{(Color online)  
Expectation value $\langle\cos\theta\rangle$ at $t=0$ as a function of $\Estatabs$ 
for
(a) $\Ialign=\SI{2e11}{\intensity}$ and 
(b) $\Ialign=\SI{e12}{\intensity}$.  
We use a $10$~ns laser pulses and $\beta=0\degree$.
The states and their labels are the same as in
  \autoref{fig:fig_4}.}
  \label{fig:fig_7}
\end{figure}

Since the energy splitting in a  pendular doublet is proportional
to the static field strength,  
the degree of adiabaticity in the 
molecular orientation should increase if $\Estatabs$ is enhanced,
\ie, the characteristic time scale of the system is reduced.
In \autoref{fig:fig_7} we present  the final orientation 
at $t=0$ of
these eight states  versus  $\Estatabs$. 
We have considered two laser pulses of $\tau=10$~ns with  peak intensities 
 $\Ialign=\SI{2e11}{\intensity}$ and  $\SI{e12}{\intensity}$, and $\beta=0\degree$. 

For the lowest laser intensity, 
the orientation of the \pstateparallel{0}{0}{e} and \pstateparallel{1}{0}{e} states 
is constant and independent of the static field for  $\Estatabs\gtrsim\SI{500}{\fieldstrength}$ 
with $|\langle\cos\theta\rangle|=0.915$. 
The orientation of the levels \pstateparallel{1}{1}{e} and \pstateparallel{2}{1}{e}  monotonically increases
as $\Estatabs$ is enhanced, and  we obtain $|\langle\cos\theta\rangle|=0.792$ for
$\Estatabs=\SI{2}{\kfieldstrength}$. 
This peak intensity is not large enough for the 
states in the third and fourth doublets to be in the
pendular regime. Thus, these pairs are weakly oriented even if
a strong dc field is used, \eg, 
for $\Estatabs=\SI{2}{\kfieldstrength}$, 
$|\langle\cos\theta\rangle|=0.265$
and $0.153$
for the third and fourth doublets, respectively.

For the strong peak intensity, all the states are in the pendular regime, and their
$|\langle\cos\theta\rangle|$ increases as $\Estatabs$ is increased reaching 
a constant value for sufficiently strong static fields.
Their orientation approaches the adiabatic limit and for 
$\Estatabs\gtrsim\SI{1}{\kfieldstrength}$, 
$|\langle\cos\theta\rangle|=0.949$ for the states of the first doublet, and 
$99.91\%$ of their population is on the corresponding adiabatic pendular state.
For $\Estatabs=\SI{2}{\kfieldstrength}$, 
the states in the fourth doublet satisfy $|\langle\cos\theta\rangle|=0.885$
and  $99.99\%$ of their population is on the corresponding adiabatic level.

In conclusion, by combining sufficiently strong electrostatic fields with 
standard Gaussian pulses, \ie,  with experimentally accessible peak intensities of
$\SI{e12}{\intensity}$ and  $10$~ns FWHM, 
a significant orientation is obtained even for excited rotational levels. 
It is worth remarking that the fields have to be parallel; then,
techniques such us the ion imaging method \cite{kupper:jcp131} could not be used to measure the
degree of orientation; whereas techniques, such as time of flight \cite{sakai:prl_90,friedrich:jmodopt50} are feasible. 


\section{Results for non-parallel fields}
\label{sec:results_beta}

In this section we investigate the rotational dynamics when the electrostatic 
field forms an angle $0\degree<\beta<90\degree$ with the  linearly polarized laser pulse. 
The azimuthal symmetry is lost, and the number of
irreducible representations is reduced to two, see \autoref{sec:hamiltonian_symmetry}. 
Thus, states with different field-free magnetic quantum numbers
are  now coupled by the electrostatic field. 

The field-free wave function of the initial state is constructed
as an  eigenstate of the operators ${\cal C}_{\Estat}(\pi)$ and $\sigma_{XZ}$, see \autoref{table:symmetry}; \ie,
\pstate{J}{M}{e}=$R_Y(\beta)$\pstate{J}{M}{e}$^0$, where $R_Y(\beta)$ is the rotation operator of an
angle $\beta$  around the LFF $Y$-axis~\cite{zare}.
This ensures that this wave function has the correct symmetries, and that 
its time evolution corresponds, in the adiabatic limit, 
to an eigenstate of the field-dressed Hamiltonian at any time.

Before the pulse is turned on, 
an important feature of the ground state 
is that its energy gap to the next state with the same symmetry is proportional
to the rotational constant $B$, which is much larger than the coupling in the weak laser field regime. 
In this regime, it evolves as an isolated state, 
and its interaction to the neighboring levels could be considered negligible. 
Hence, analogously to the parallel field configuration, the formation of the doublets in the pendular
regime is the only source of non-adiabatic effects in its  field-dressed dynamics. 
Note that the lowest lying level of the odd irreducible representation will 
show the same behavior.

\begin{figure}[htb]
  \centering
  \includegraphics[width=.48\textwidth]{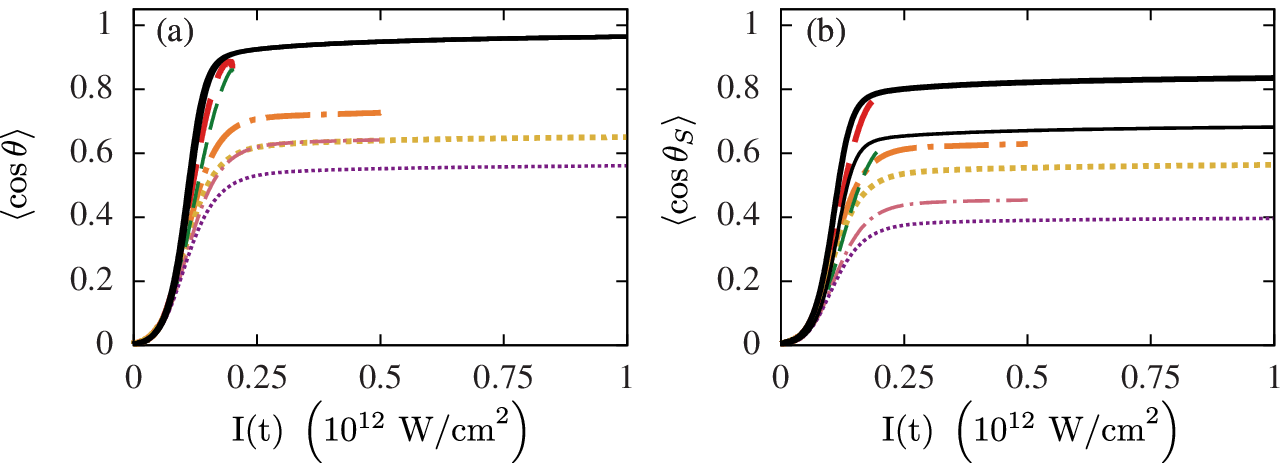}
\caption{(Color online) For the ground state, evolution of the expectation values
(a) $\langle\cos\theta\rangle$ 
and 
(b) $\langle\cos\theta_{\textup{s}}\rangle$ 
as a function of $\textup{I}(t)$ of a $10$~ns Gaussian pulse.
The field configurations are 
$\beta=30\degree$ (thick) and $45\degree$ (thin) with peak intensities 
$\Ialign=\SI{2e11}{\intensity}$ (dashed), 
$\SI{5e11}{\intensity}$ (dot-dashed),  and
$\SI{e12}{\intensity}$ (dotted). 
The static electric field is fixed to $\Estatabs=\SI{300}{\fieldstrength}$. 
The adiabatic results (solid) are also included.}
  \label{fig:fig_8}
\end{figure}

For the ground state, the time evolution of the expectation values
 $\langle\cos\theta\rangle$ and  $\langle\cos\theta_{\textup{s}}\rangle$ 
 are presented as
a function of $\textup{I}(t)$ till the peak intensity is reached in 
\autoref{fig:fig_8}~(a) and~(b). 
The Gaussian pulse has $10$~ns FWHM and peak intensities 
$\Ialign=\SI{2e11}{\intensity}$, 
$\SI{5e11}{\intensity}$,  and
$\SI{e12}{\intensity}$. 
We consider the inclination angles 
$\beta=30\degree$ and $45\degree$, and a dc field of
$\Estatabs=\SI{300}{\fieldstrength}$.
For comparison, the adiabatic results are also 
included,  with $\langle\cos\theta\rangle$ being independent of  $\beta$. 
For a certain laser pulse, increasing the inclination angle towards $90\degree$ implies 
a decrease of the energy splitting in the first doublet, $\Delta E\sim2\mu\Estatabs\cos\beta$,
 and, therefore, 
an increase of the adiabatic parameter \eqref{eq:adiabatic_criteria}. 
Note that in the energy splitting we have not considered the
component of the dc field  along the LFF $X$-axis. 
Compared to the $\beta=0\degree$ configuration, 
the dynamics could be considered as less adiabatic being characterized by a larger
population transfer to the other adiabatic state in this pendular pair.
The final orientation is significantly decreased as $\beta$ is increased,
\eg, for $\Ialign=\SI{e12}{\intensity}$,  
$\langle\cos\theta\rangle=0.651$ and $0.561$, 
and the contribution of the adiabatic ground state is
$|C_{00e}(0)|^2=0.837$ and  $0.791$ for 
$\beta=30\degree$ and  $45\degree$, respectively. 
For a certain angle $\beta$, the orientation achieved at $t=0$ decreases as $\Ialign$ is
increased, cf. \autoref{fig:fig_8}.
Since the molecular dynamics of the ground state is restricted to the two lower pendular adiabatic
states, its time dependent results for its energy and expectation values $\langle\cos^2\theta\rangle$ and 
$\langle\mathbf{J}^2\rangle$ resemble the adiabatic ones. 

\begin{figure}[htb]
  \centering
  \includegraphics[width=.45\textwidth]{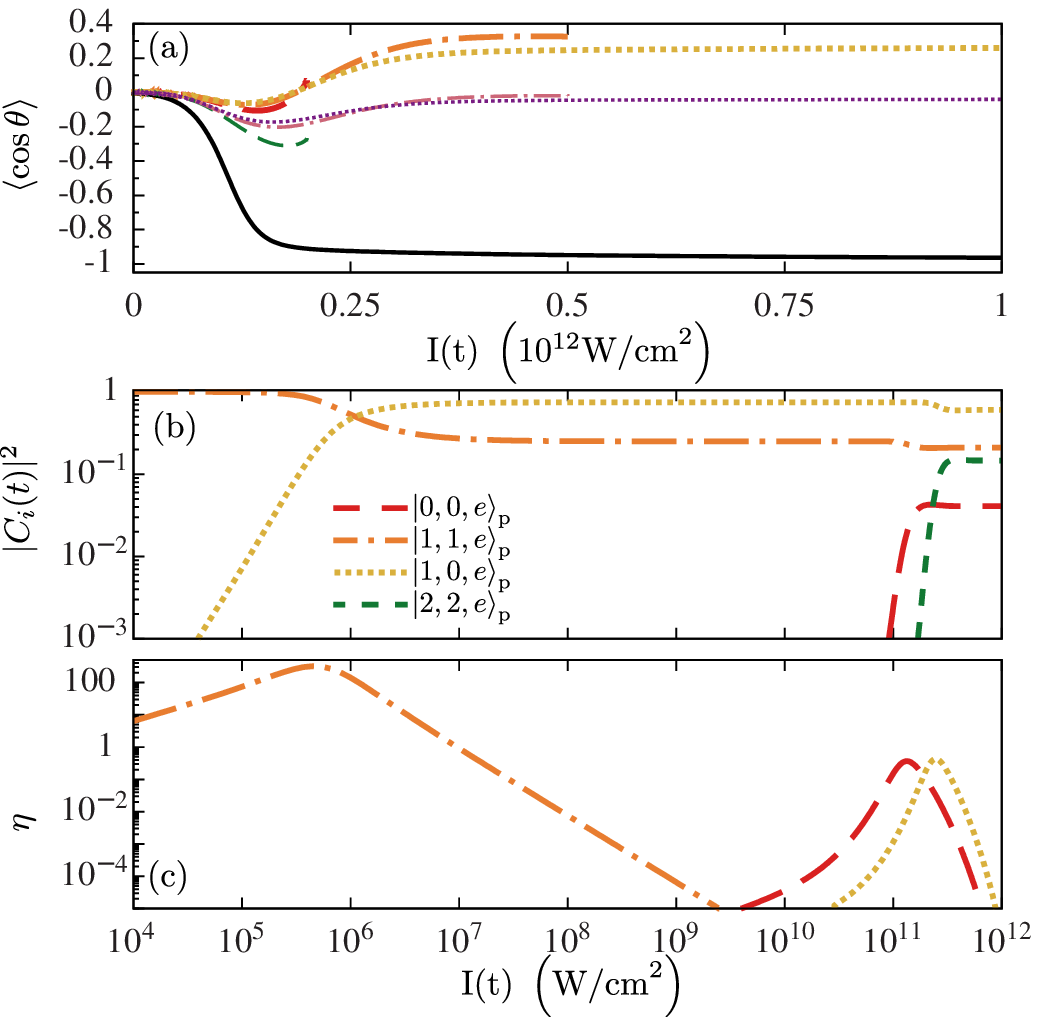}
  \caption{(Color online) 
(a) For the \pstate{1}{1}{e} state, expectation value
$\langle\cos\theta\rangle$ as a function of $\textup{I}(t)$ of a $10$~ns Gaussian pulse.
The field configurations are $\beta=30\degree$ (thick) and $45\degree$ (thin), with
$\Ialign=\SI{2e11}{\intensity}$ (dashed), 
$\SI{5e11}{\intensity}$  (dot-dashed) and
$\SI{e12}{\intensity}$ (dotted).
The adiabatic results (solid line) are also included.
For $\Ialign=\SI{e12}{\intensity}$ and  $\beta=30\degree$, (b) 
square of the projections of 
the time-dependent wave function on the adiabatic pendular states
\ppstate{1}{1}{e} (dot-dashed), 
\ppstate{1}{0}{e} (dotted), 
\ppstate{0}{0}{e} (long-dashed), and  
\ppstate{2}{2}{e} (short-dashed), and 
(c) 
adiabatic parameter  between the pendular states
\ppstate{1}{1}{e} and \ppstate{1}{0}{e} (dot-dashed), 
\ppstate{1}{1}{e} and \ppstate{0}{0}{e} (dashed),  and 
\ppstate{1}{0}{e} and \ppstate{2}{2}{e} (dotted).
The dc field is fixed to $\Estatabs=\SI{300}{\fieldstrength}$. }
  \label{fig:fig_9}
\end{figure}

To illustrate the rotational  dynamics of excited states, 
we show in \autoref{fig:fig_9}~(a) 
the orientation cosine  $\langle\cos\theta\rangle$ 
as a function of $\textup{I}(t)$ for \pstate{1}{1}{e}. 
The field configurations are the same as in 
\autoref{fig:fig_8}. 
The adiabatic model predicts a sharp wrong-way orientation. In contrast, 
this state presents a weak or even zero orientation, and 
in addition, a larger peak intensity does not imply a larger orientation.
When the peak intensity is reached, this level shows a weak right-way orientation for
$\beta=30\degree$: $\langle\cos\theta\rangle=0.326$   and
$0.259$, for $\Ialign=\SI{5e11}{\intensity}$  and
$\SI{e12}{\intensity}$, respectively. 
For $\beta=45\degree$ and the peak intensities $\Ialign=\SI{5e11}{\intensity}$  and
$\SI{e12}{\intensity}$,  due to the non-adiabatic dynamics 
\pstate{1}{1}{e} is not oriented.

Let us analyze in detail these results. 
For  highly excited states, the dynamics is more complicated. 
Apart from the doublet formation, there is another physical phenomenon 
at weak laser intensities which causes loss of adiabaticity. 
In the presence of only a weak static field, 
the $M$-degeneracy of the states with the same 
field-free $J$ is broken due to the  quadratic Stark effect, 
\ie, the splitting goes as $\Delta E\sim \Estatabs^2$.
As the pulse is switched on, the energy gap between two states of this $J$-manifold  is much smaller than the rate of
their coupling due to the laser field,  
\ie, $\eta$ is larger than one. 
For  $\beta=30\degree$, 
the adiabatic parameter $\eta$ between the states
\ppstate{1}{1}{e} and \ppstate{1}{0}{e}, both contributing to the dynamics of \pstate{1}{1}{e}, 
is presented in  \autoref{fig:fig_9}~(b), and it 
achieves large values for $\textup{I}(t)\lesssim\SI{5e5}{\intensity}$.
As the states in this  $J$-manifold are driven apart by the laser field, the process is
non-adiabatic and there is a population transfer between them. 
The  projections of the time-dependent 
wave function in terms
of the adiabatic states
\ppstate{0}{0}{e}, 
\ppstate{1}{1}{e}, 
\ppstate{1}{0}{e}, 
and
\ppstate{2}{2}{e} 
is  presented in 
\autoref{fig:fig_9}~(d) for $\beta=30\degree$ and
$\Ialign=\SI{e12}{\intensity}$.
Under these diabatic conditions, $|C_{11e}(t)|^2$ decreases as $\textup{I}(t)$ is increased, whereas  $|C_{10e}(t)|^2$ increases.
By further increasing $\textup{I}(t)$, 
the coupling between these states becomes very small or even zero and their energy separation
increases, so that  
$\eta$ decreases and the population transfer is stopped.
This process is so diabatic that the wave function does not change, but its projections on the
adiabatic basis are modified because the basis varies with time.
For instance, the field-free state is 
\pstate{1}{1}{e}$=\cos\beta$\ppstateparallel{1}{1}{e}$+\sin\beta$\ppstateparallel{1}{0}{e},
which belongs to the proper irreducible representation.
After swichting on the static field, its wave function 
could be approximated by the same expression because this field is very weak.  
Once the splitting of this $J$-manifold is finished, \ie,  for $\textup{I}(t)\sim \SI{5e7}{\intensity}$,  
the contributions of the states \ppstate{1}{1}{e} and \ppstate{1}{0}{e} are approximately 
 $\sin^2\beta$ and $\cos^2\beta$, respectively. Note that the states \ppstateparallel{J}{M}{e} and
\ppstate{J}{M}{e} are not related adiabatically.

In contrast  to the ground state,
the wave function of any excited level has contributions from 
adiabatic states which correspond to different pendular doublets. 
As the laser intensity is increased, the  molecular dynamics is affected by
the formation of these pendular doublets. Thus,  
the final orientation could be significantly reduced compared to the parallel fields result. 
For instance, the time-dependent \pstate{1}{1}{e} state
has contributions from the adiabatic  levels \ppstate{1}{1}{e} and \ppstate{1}{0}{e}, which
correspond to the first and second pendular doublets, respectively. 
In \autoref{fig:fig_9}~(b)
we show how the adiabatic parameters  $\eta$
between the pairs \ppstate{0}{0}{e}-\ppstate{1}{1}{e} and \ppstate{1}{0}{e}-\ppstate{2}{2}{e}, 
which form the first and second doublets, respectively, achieve  values close to $1$. 
The final population of the state   \pstate{1}{1}{e} is 
$|C_{00e}(0)|^2=0.041$,  $|C_{11e}(0)|^2=0.210$, $|C_{10e}(0)|^2=0.603$, and  $|C_{22e}(0)|^2=0.146$, which
gives rise to a small orientation
As a consequence of this population redistribution to other pendular doublets, 
features of the system such as energy, alignment and hybridization of the angular motion
do not resemble the adiabatic results. In particular, since 
the levels on the second pendular doublet possess a smaller alignment,
the adiabatic result is larger than the time-dependent one.
For instance, for $\beta=30\degree$ and $\Ialign=\SI{e12}{\intensity}$, 
once the time evolution is finished 
the alignment of this state \pstate{1}{1}{e} is $\langle\cos^2\theta\rangle=0.879$, 
compared to $\langle\cos^2\theta\rangle= 0.931$ 
obtained for the adiabatic level \ppstate{1}{1}{e}.

For $\beta=45\degree$, despite the fact that  the \pstate{1}{1}{e} level is significantly aligned, 
$\langle\cos^2\theta\rangle=0.896$, 
it is not  oriented with 
$\langle\cos\theta\rangle=-0.041$ for $\Ialign=\SI{e12}{\intensity}$. 
This state does not gain any orientation 
if stronger peak intensities are used. 
This is a consequence of the population redistribution explained above. Indeed,  this level
could be considered as a {\it dark state} for the mixed-field orientation dynamics.  
This physical phenomenon is not restricted to this state
and field configuration. We show below that other levels also  behave as {\it dark states}. 
It is worth noting that if in a mixed-field orientation experiment 
these dark states form part of the molecular
beam, 
the degree of orientation will be diminished.

The population redistribution to other pendular doublets significantly affects  
the expectation value $\langle\cos\theta_{\textup{s}}\rangle$. 
To $\langle\cos\theta_{\textup{s}}\rangle$ contribute
terms which mix up adiabatic states
with different magnetic quantum numbers 
Since their wave functions could spatially overlap, their coupling matrix elements
do not vanish, and $\langle\cos\theta_{\textup{s}}\rangle$ oscillates as $t$ is
increased. 

The phenomenon of population redistribution 
at weak laser intensities also occurs for highly excited rotational levels, and for them,
more adiabatic states would be involved in it. 
Before the Gaussian pulse is turned on, 
the Stark separation of the states  in a certain $J$-manifold
is increased if 
the electrostatic field strength is enhanced.  
Then, the adiabatic parameter $\eta$ is reduced, and 
the process of splitting of this $J$-manifold  becomes less diabatic. 
Indeed, for sufficiently strong dc-fields, 
the dynamics would be adiabatic 
without population transfer between the states with the same field-free $J$. 
For instance, the mixed-field dynamics of the  \pstate{1}{1}{e}   level
can be considered as adiabatic on the weak laser field regime
for
$\Estatabs\gtrsim\SI{14}{\kfieldstrength}$ and $\beta=30\degree$.

Let us remark that the excited states could also suffer 
avoided crossings with adjacent levels, having 
different field-free magnetic quantum number $M$, 
as the pulse intensity is varied.
The rotational dynamics in most of these crossings will be non-adiabatic~\cite{omiste:pccp2011}. 

\subsection{Influence of the peak intensity $\Ialign$}
\label{sec:intensity_beta}

\begin{figure}[h]
  \centering
  \includegraphics[width=.5\textwidth]{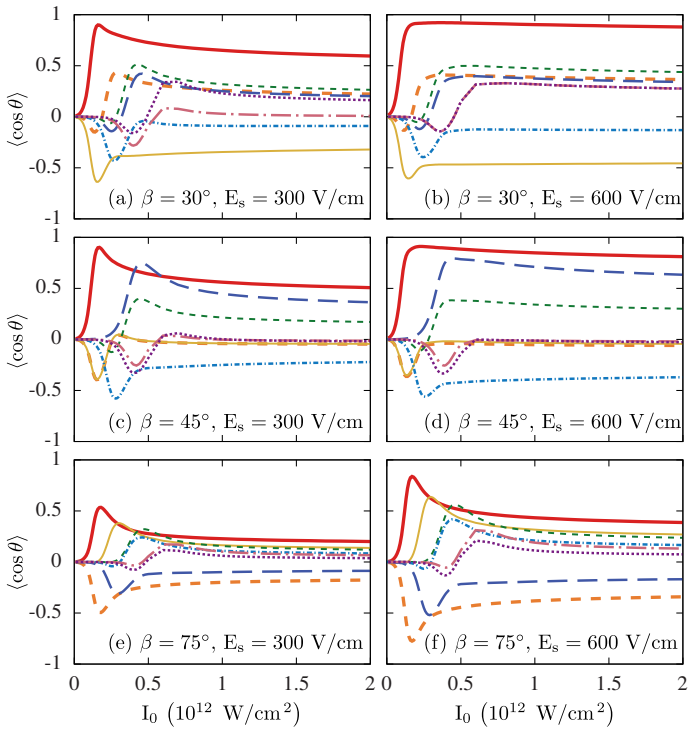}
  \caption{(Color online)   
Expectation value $\langle\cos\theta\rangle$ at $t=0$ as a function of 
the peak intensity $\Ialign$ for 
\pstate{0}{0}{e} (red thick solid), 
\pstate{1}{0}{e} (gold thin solid)
\pstate{1}{1}{e} (orange thick short-dashed),
\pstate{2}{1}{e} (dark blue long-dashed)
\pstate{2}{0}{e} (blue dot-short-dashed), 
\pstate{3}{0}{e} (purple dotted),
\pstate{2}{2}{e} (green thin short-dashed) and
\pstate{3}{2}{e} (pink dot-long-dashed).
The field configurations are 
(a)-(b) $\beta=30\degree$, 
(c)-(d) $\beta=45\degree$ and
(e)-(f) $\beta=75\degree$, 
with $\Estatabs=\SI{300}{\fieldstrength}$ and $\SI{600}{\fieldstrength}$, respectively. 
The FWHMs of the Gaussian pulses are fixed to $10$~ns.}
  \label{fig:fig_10}
\end{figure}

Analogously to the parallel-field configuration, we investigate now the impact of the
laser peak intensity on the orientation. 
To do so, we restrict this study to the following 
eight states: 
\pstate{0}{0}{e},  
\pstate{1}{0}{e}, 
\pstate{1}{1}{e}, 
\pstate{2}{0}{e}, 
\pstate{2}{1}{e}, 
\pstate{2}{2}{e}, 
\pstate{3}{0}{e}, 
and 
\pstate{3}{2}{e}. 
Note that they are related to the ones analyzed in the parallel fields configuration, by  a
rotation of $\beta$ around the LFF $Y$ axis. 
Their orientation at $t=0$, \ie,  $\langle\cos\theta\rangle$
for $\textup{I}(0)=\Ialign$, is plotted as a function of $\Ialign$ in  \autoref{fig:fig_10}
for $\beta=30\degree$, panels~(a)-(b), $\beta=45\degree$, panels~(c)-(d), 
and $\beta=75\degree$, panels~(e)-(f), and 
$\Estatabs=\SI{300}{\fieldstrength}$ and $\SI{600}{\fieldstrength}$, respectively.
The FWHM of these Gaussian  pulses is fixed to  $\tau=10$~ns. 

Let us start analyzing the results for the ground state. 
For all field configurations, $\langle\cos\theta\rangle$ shows a 
qualitatively similar behavior as a function of the peak intensity: 
initially increases, reaches a maximum and decreases thereafter.
At the peak intensity where the maximum of $\langle\cos\theta\rangle$  takes place, the dynamics of
this state is non-adiabatic, but due to the coupling of both states the orientation increases with respect
to the adiabatic result.
For a fixed peak intensity and electric field strength,  $\langle\cos\theta\rangle$ decreases
as $\beta$ is increased towards $90\degree$. 
For $\beta=75\degree$ and $\Ialign= \SI{2e11}{\intensity}$, the ground state 
achieves a moderate maximal orientation, $\langle\cos\theta\rangle=0.514$  
and $0.796$ for $\Estatabs=\SI{300}{\fieldstrength}$ and $\SI{600}{\fieldstrength}$, respectively.

The population transfer taking place at weak and strong laser intensities leaves its
finger-prints in the dynamics of the excited states.
Compared to the parallel field results, cf \autoref{fig:fig_4},
their orientation is reduced for any inclination angle $\beta$ and  
the pendular pairs are not any longer formed  by  a right- and wrong-way oriented states.
Whereas for most of the field configurations, the ground state possesses the largest orientation,
the levels \pstate{1}{0}{e} or  \pstate{2}{1}{0} could achieve a similar or even larger orientation,
\eg, 
for $\beta=45\degree$ and $ 75\degree$, $\Ialign\approx \SI{5e11}{\intensity}$ and $\Estatabs=\SI{300}{\fieldstrength}$.
For $\beta=30\degree$  $\Estatabs=\SI{600}{\fieldstrength}$,
 the degree of orientation is moderate for most of the states. 
Several dark states are found for  $\beta=45\degree$:
\pstate{1}{1}{e},
\pstate{1}{0}{e}, 
\pstate{3}{0}{e},
and 
\pstate{3}{2}{e}, see \autoref{fig:fig_4}~(c)-(d).
For instance,
the levels \pstate{1}{1}{e} and 
\pstate{1}{0}{e} 
are strongly aligned with $\langle\cos^2\theta\rangle=0.927$
for $\Ialign\gtrsim \SI{2e12}{\intensity}$ and
$\Estatabs=\SI{600}{\fieldstrength}$, 
whereas they are not orientated with 
$\langle\cos\theta\rangle\approx-0.059$ and $-0.043$, respectively.  
For $\beta=75\degree$, when the peak intensity of the Gaussian pulse is reached 
most of the states present a weak orientation,
only the  levels \pstate{0}{0}{e} and \pstate{1}{1}{e}
have a large orientation for small values of  $\Ialign$. 

These results indicate that with a $10$~ns alignment pulse, strong dc fields and
small inclination angles are required to reach a moderate orientation for excited states.

\subsection{Influence of the FWHM of the laser pulse}
\label{sec:width_beta}
\begin{figure}[h]
  \centering
  \includegraphics[width=.5\textwidth]{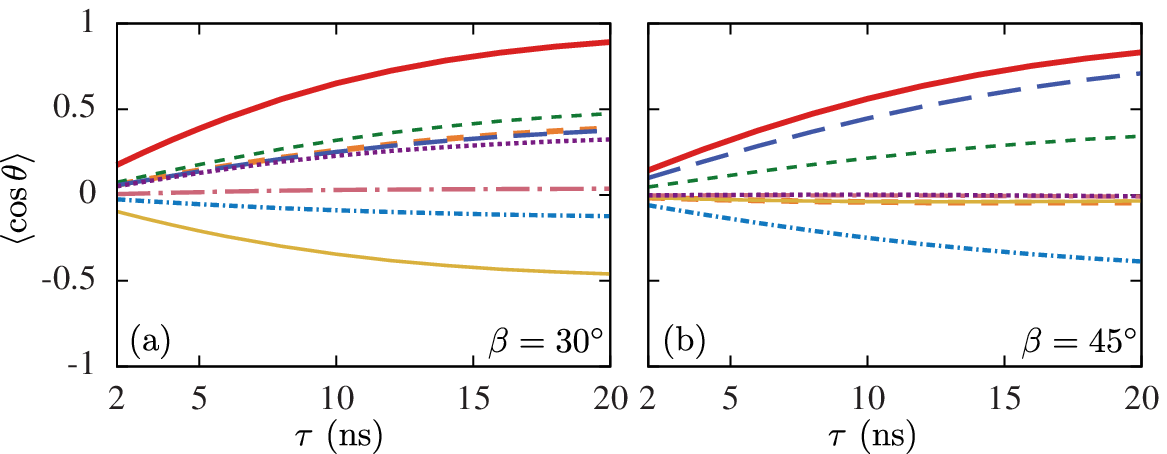}
\caption{(Color online) Expectation value $\langle\cos\theta\rangle$   at $t=0$   as a function of 
$\tau$. The field configurations are 
(a) $\beta=30\degree$ 
and  
(b) $\beta=45\degree$, with $\Ialign=\SI{e12}{\intensity}$ and $\Estatabs=\SI{300}{\fieldstrength}$. 
The states and their labels are the same as in \autoref{fig:fig_10}.}
\label{fig:fig_11}
\end{figure}

For the same set of states as in the previous section,
we analyze here how their directional properties depend on the
FWHM of the Gaussian pulse.
In \autoref{fig:fig_11}~(a) and (b)
we show $\langle\cos\theta\rangle$  at $t=0$  as 
a function of $\tau$ for $\beta=30\degree$ and $45\degree$, respectively.
The static electric field is fixed to $\Estatabs=\SI{300}{\fieldstrength}$,
and the peak intensity to
$\Ialign=\SI{e12}{\intensity}$.

The orientation of the ground state increases approaching the adiabatic limit as $\tau$ is
increased, and it reaches it with  a $50$~ns pulse.
We encounter several excited states presenting a moderate or weak orientation, and
their  $|\langle\cos\theta\rangle|$ monotonically increases as $\tau$ is enhanced,
\eg, for $\beta=30\degree$ the levels \pstate{1}{0}{e}, \pstate{1}{1}{e},
\pstate{2}{0}{e}, \pstate{2}{1}{e}, \pstate{2}{2}{e} and \pstate{3}{0}{e}
and for $\beta=45\degree$ 
\pstate{2}{0}{e}, \pstate{2}{1}{e} and \pstate{2}{2}{e}.
For all of them, a $20$~ns pulse is not enough to achieve the adiabatic regime.
In contrast, other excited levels present a very small, almost zero, orientation 
independently of the pulse duration.
Some of these levels behave as dark states
being strongly aligned but not oriented independently of the pulse duration,
\eg, the \pstate{3}{2}{e} state
has  $\langle\cos^2\theta\rangle=0.755$ and $|\langle\cos\theta\rangle|<0.04$
for  $\beta= 30\degree$ and any value of $\tau$.
An analogous behavior is found for the levels \pstate{1}{0}{e}, \pstate{1}{1}{e}, \pstate{3}{0}{e} and \pstate{3}{2}{e}
at $\beta=45\degree$. 
As described above, this phenomenon is due to the non-adiabatic dynamics at
weak laser intensities when the levels of the $J$-manifold are driven apart,
and it takes places even for  $50$~ns pulses.

\subsection{Influence of the electrostatic field strength}
\label{sec:static_beta}

\begin{figure}[h]
  \centering
  \includegraphics[width=.5\textwidth]{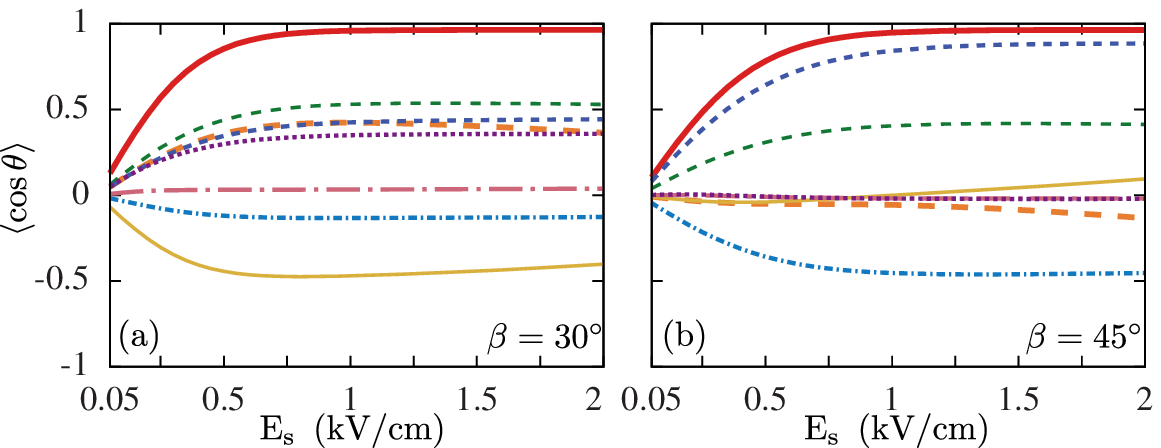}
  \caption{(Color online)  
Expectation value of $\langle\cos\theta\rangle$ at $t=0$  as a function of $\Estatabs$
for the field configurations $\tau=10$~ns, $\Ialign=\SI{e12}{\intensity}$, and 
(a) $\beta=30\degree$ and 
(b) $\beta=45\degree$. 
The labeling of the states is done as in \autoref{fig:fig_10}.}
  \label{fig:fig_12}
\end{figure}

In this section, we consider two inclination angles, and investigate the impact of the electrostatic
field on the mixed-field orientation dynamics of the 
same states.
Figures \ref{fig:fig_12}~(a) and (b) illustrate 
the behavior of $\langle\cos\theta\rangle$ at $t=0$ as a function of $\Estatabs$ 
 for $\beta=30\degree$ and $45\degree$, respectively.
The laser pulse has $\tau=10$~ns and $\Ialign=\SI{e12}{\intensity}$. 

The ground state presents the largest orientation, which increases as $\Estatabs$ is enhanced,
being strongly oriented for sufficiently large fields, 
\eg,  $\langle\cos\theta\rangle>0.9$ for $\Estatabs\ge\SI{600}{\fieldstrength}$ and
$\beta=30\degree$.   
Regarding the excited states, their orientation strongly depends on the inclination angle.
For $\beta=30\degree$, $|\langle\cos\theta\rangle|$ 
monotonically increases till it reaches a plateau-like behavior, and they 
show a moderate orientation. Indeed, for $\beta=30\degree$, $\Ialign=\SI{e12}{\intensity}$ and  
$\Estatabs=\SI{2}{\kfieldstrength}$,
we obtain at the maximum of the Gaussian pulse 
$\langle\cos\theta\rangle=-0.402$ and $0.365$
for the states  \pstate{1}{0}{e}  and \pstate{1}{1}{e}, respectively.  
For $\beta=45\degree$, the level \pstate{2}{1}{e} presents a large orientation: $\langle\cos\theta\rangle>0.8$
for $\Estatabs\gtrsim\SI{800}{\fieldstrength}$.  
There are some darks states for $\beta=45\degree$, which are not oriented even when  dc fields of $\SI{2}{\kfieldstrength}$ are
used, \eg,  \pstate{3}{0}{e} and \pstate{3}{2}{e}. 

For non-parallel fields, a strong dc field does not 
ensure a large orientation for  excited rotational states. 
If the aim is a strongly oriented molecular ensemble, then this should 
be as pure as possible in the ground state.

In the Hamiltonian \eqref{eq:hs}, the term $-\mu \Estatabs\sin\beta\sin\theta\cos\phi$
is responsible for the mixing of states 
with different  field-free magnetic quantum numbers. 
On the weak dc field regime, 
the mixing between these states is so small that $M$ could be considered as conserved, and
this term could be neglected. 
By increasing $\Estatabs$, this coupling between levels with different 
field-free $M$ becomes important, and 
this should affect the molecular dynamics. 
Thus, the questions that remain open is how important is the $X$-component of the electrostatic field
to the dynamics, and for which electric field regime, we
could only consider
its $Z$-component $\mathbf{E}_{\textup{s}}=\Estatabs\cos\beta \hat{Z}$.

As indicated above, even for tilted fields, the dynamics of the ground state can be described by  a
two state 
model. Its energy separation to the next state with $M\ne0$ is of the order of $B$ and larger 
than the dc field coupling to these levels. 
Thus, for $\Estatabs\lesssim\SI{20}{\kfieldstrength}$, the dynamics considering 
the dc field is equal to the one obtained when
only its $Z$-component is included. 

For the excited states, 
the answer to these questions depends on how the initial wave function, before the fields are
switched on, is constructed. 
The first option is to proceed as indicated at the beginning of this section; 
the field-free  $\beta\ne0\degree $ and $\beta=0\degree$ wave functions are related 
by a rotation of $\beta$ around the $Y$-axis \pstate{J}{M}{e}=$R_Y(\beta)$\pstate{J}{M}{e}$^0$.  
In this case,  for  the level \pstate{1}{1}{e},
 some differences  in its orientation are observed for $\Estatabs\gtrsim\SI{1}{\kfieldstrength}$
 with  $|\langle\cos\theta\rangle|$ being  larger if the two components of $\mathbf{E}_{\textup{s}}$  are considered.
These differences are augmented as $\Estatabs$ is increased,
\eg, for a $10$~ns laser pulse with  $\Ialign= \SI{1e12}{\intensity}$, $\beta=45\degree$ and
$\Estatabs=\SI{5}{\kfieldstrength}$, we obtain at $t=0$ 
$\langle\cos\theta\rangle=-0.629$ compared to
$\langle\cos\theta\rangle=-0.019$ if only the $Z$-component of $\mathbf{E}_{\textup{s}}$ is included.
By increasing $\Estatabs$ this state will achieve an adiabatic dynamics only if
both components of the static field are present.
The second option is to construct  the field-free $\beta\ne0\degree $ wave function equal to the 
field-free $\beta=0\degree $ one. 
In this case, the results resemble those of the parallel field configuration taking into
account  $\cos\beta$ as scaling factor for the static field strength.


\subsection{Influence of the inclination of the fields} 
\label{sec:beta}

The symmetries of the rotational Hamiltonian \eqref{eq:hamiltonian},
see \autoref{sec:hamiltonian_symmetry},
 and, therefore,  the rotational dynamics  strongly depend on 
the angle between  the fields. 
In this section, we investigate in detail the impact of the inclination angle in 
the mixed-field orientation dynamics.

\begin{figure}[h]
  \centering
  \includegraphics[width=.45\textwidth,angle=0]{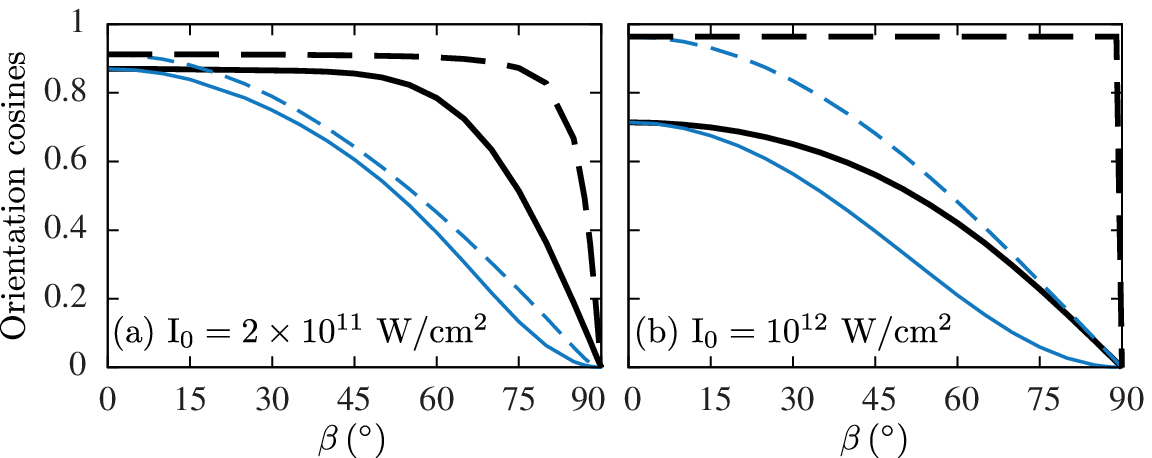}
  \caption{(Color online) Expectation values $\langle\cos\theta\rangle$ (thick solid line) 
and $\langle\cos\theta_{\textup{s}}\rangle$ (thin solid line) at $t=0$ 
as a function of $\beta$
for the ground state.  
The peak intensities are
(a) $\Ialign= \SI{2e11}{\intensity}$  and 
(b) $\Ialign= \SI{e12}{\intensity}$. 
The adiabatic results for $\langle\cos\theta\rangle$ (thick dashed line) 
and $\langle\cos\theta_{\textup{s}}\rangle$ (thin dashed line)
are also presented. 
The FWHM of the laser pulse is fixed to $\tau=10$~ns and the dc field to $\Estatabs=\SI{300}{\fieldstrength}$.}
  \label{fig:fig_13}
\end{figure}

 For the ground state  \pstate{0}{0}{e}, 
the orientation cosines  $\langle\cos\theta\rangle$ and $\langle\cos\theta_{\textup{s}}\rangle$
are plotted in \autoref{fig:fig_13}, 
as a function of $\beta$, together with the adiabatic results. 
For a weak dc field and strong laser field, the following relation
$\langle\cos\theta_{\textup{s}}\rangle\approx \langle\cos\theta\rangle\cos \beta$ is satisfied 
within  the adiabatic limit. 
In  $\langle\cos\theta_{\textup{s}}\rangle$ 
the term $\langle\sin\theta\cos\phi\rangle\sin\beta$ has been
neglected, which can be done as far as the mixing between states with different  
field-free $M$ is very small. By increasing the electrostatic field strength, a
regime would be encountered  where this approximation does not hold any longer.
An analogous  relation is satisfied between the time-dependent orientation cosines of the ground state. 
For $\Ialign=\SI{2e11}{\intensity}$, its orientation  $\langle\cos\theta\rangle$ 
shows a plateau-like behavior 
till $\beta=50\degree$, which is very close to the adiabatic limit.
By further increasing $\beta$, 
$\langle\cos\theta\rangle$ decreases and approaches to zero.
For $\beta=90\degree$, the states in a pendular doublet have different symmetry 
and are not coupled by the dc field, thus they might be strongly aligned 
but not oriented.
For $\Ialign=\SI{e12}{\intensity}$, $\langle\cos\theta\rangle$ 
monotonically decreases as $\beta$ is increased towards $90^{\circ}$,
and its value is always smaller than for $\Ialign=\SI{2e11}{\intensity}$. 
For both laser fields, $\langle\cos\theta_\textup{s}\rangle$ decreases as $\beta$ is increased.

\begin{figure}[h]
  \centering
  \includegraphics[width=.48\textwidth,angle=0]{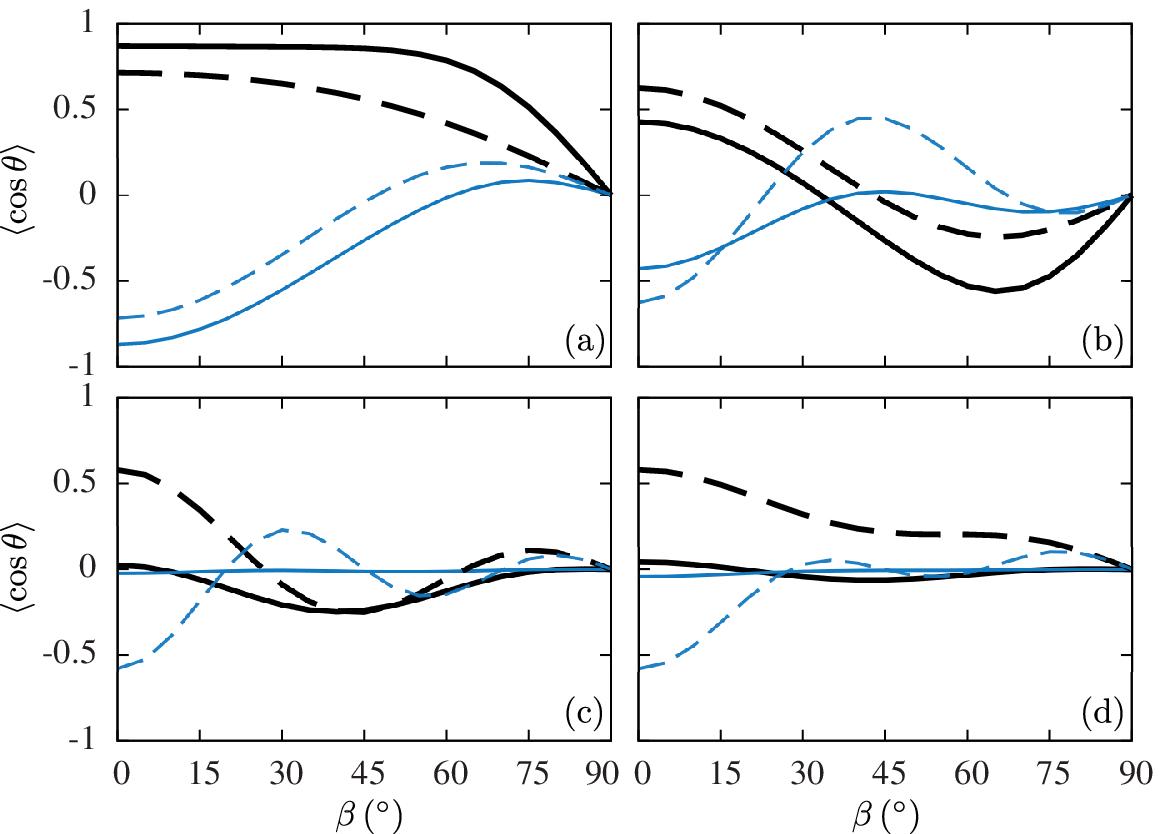}
  \caption{(Color online) Expectation value $\langle\cos\theta\rangle$ at $t=0$ as a function of $\beta$
for the states
(a) \pstate{0}{0}{e} (black), 
and  
\pstate{1}{0}{e} (blue), 
(b)
\pstate{1}{1}{e} (black),
and
\pstate{2}{1}{e} (blue), 
(c)
\pstate{2}{0}{e} (black),
and
\pstate{3}{0}{e} (blue), 
(d)
\pstate{2}{2}{e} (black), 
and 
\pstate{3}{2}{e} (blue).  
The field configuration is $\Estatabs=\SI{300}{\fieldstrength}$ and 
$\Ialign= \SI{2e11}{\intensity}$ (solid lines) and
$\SI{e12}{\intensity}$ (dashed lines).
The FWHM of the laser pulse is fixed to $\tau=10$~ns. }
  \label{fig:fig_14}
\end{figure}

In Figs.~\ref{fig:fig_14}~(a), (b), (c) and  (d), 
we present  the  orientation cosine of the pairs
 \pstate{0}{0}{e}-\pstate{1}{0}{e}, 
\pstate{1}{1}{e}-\pstate{2}{1}{e}, 
\pstate{2}{2}{e}-\pstate{3}{2}{e} and 
\pstate{2}{0}{e}-\pstate{3}{0}{e},  respectively, as a function  $\beta$. 
 The static field strength is 
$\Estatabs=\SI{300}{\fieldstrength}$ and we
consider two Gaussian pulses of $10$~ns FWHM and peak intensities
$\Ialign= \SI{2e11}{\intensity}$ and $\SI{e12}{\intensity}$. 
Due to the complicated field-dressed dynamics of excited states for $0\degree<\beta<90\degree$ with contributions from
several pendular pairs, 
in $\langle\cos\theta_{\textup{s}}\rangle$ the term $\langle\sin\theta\cos\phi\rangle$
cannot be neglected. 
Then, the simple relation
$\langle\cos\theta_{\textup{s}}\rangle\approx\langle\cos\theta\rangle\cos\beta$ 
does not hold for these levels.

Based on the adiabatic theory, 
the ground state and the level \pstate{1}{0}{e}, 
should present the same orientation but with opposite directions. 
However, this is only satisfied for $\beta=0\degree$.
Due to the non-adiabatic effects at weak laser intensities, 
its $|\langle\cos\theta\rangle|$ is
smaller than the corresponding value of \pstate{0}{0}{e}  for $0<\beta<90\degree$.
For the second doublet, cf. \autoref{fig:fig_14}~(b),  
$\langle\cos\theta\rangle$ oscillates as $\beta$ is varied, and the orientation even changes its
direction. 
Both states could present a moderate orientation at a certain value of $\beta$. 
The pendular regime is not achieved by the third and fourth pairs with a 
$10$~ns laser 
pulse and  $\Ialign=\SI{2e11}{\intensity}$,
and their orientation is either zero or very small independently of $\beta$. 
For $\Ialign=\SI{e12}{\intensity}$ and $\beta=0\degree$, these four states show a moderate 
 orientation, which is reduced for any other angle, being small for
$\beta\gtrsim 60\degree$. 
At the strong peak intensity $\Ialign=\SI{e12}{\intensity}$, 
in all pendular doublets one of the two
levels  presents the dark behavior with respect to the mixed-field orientation dynamics at
a certain angle $\beta$. 

These results show that if the molecular beam is rotationally cold,
a small inclination angle will optimize the degree of orientation observed in the experiment. 


\section{Rotational dynamics once the laser pulse is turned on}
\label{sec:dynamics_top_pulse}

Let us investigate the dynamics for $t>0$ assuming that 
the laser peak intensity, reached at $t=0$, and the dc field strength are kept constant for $t>0$; 
\ie, $\textup{I}(t)=\Ialign$ and $\textup{E}_\text{s}(t)=\textup{E}_\textup{s}$ for  $t>0$.
At $t=0$, the time-dependent wave function can be expressed in terms
of the corresponding adiabatic basis. Since the Hamiltonian is
time-independent for $t>0$, the contribution of each adiabatic state remains constant as $t$ is increased. 
For a certain state $|\gamma\rangle$, the expectation value of an operator $\hat{A}$ in this
adiabatic basis reads as 
  \begin{eqnarray}
  \label{eq:exp_value}
    \langle\hat{A}\rangle&=&
    \sum_{j}\left|C_j(0)\right|^2\tensor[_{\textup{p}}]{\langle \gamma_j|\hat{A}|\gamma_j\rangle}{_{\textup{p}}}\\
\label{eqn:oscillate_cosine}
&+&2\sum_{j<k}|C_j(0)||C_k(0)|\tensor[_{\textup{p}}]{\langle\gamma_j|\hat{A}|\gamma_k\rangle}{_{\textup{p}}}\cos\left(\cfrac{\Delta
        E_{jk}t}{\hbar}+\delta_{jk}\right)     \nonumber,  
  \end{eqnarray}
with $C_j(0)$ being the weight   at $t=0$ of the adiabatic state $|\gamma_j\rangle_\textup{p}$ 
to the wave function  of $|\gamma\rangle$,
$\Delta E_{jk}$ the energy splitting between the adiabatic levels
$|\gamma_j\rangle_\textup{p}$ and $|\gamma_k\rangle_\textup{p}$, and $\delta_{jk}$ the phase difference of 
$C_j(0)$ and $C_k(0)$. 

Based on the results presented above, 
the time-dependent wave function could have contributions from: 
i) only the adiabatic levels forming a pendular doublet,
or ii) several adiabatic levels from at least two  pendular doublets.
All the states for $\beta=0\degree$, and the ground
state for $0\degree\le\beta<90\degree$ could belong to the first case. Whereas, the second one
refers to all excited states when $0\degree<\beta<90\degree$, unless the static field is very strong. 

\begin{figure}[h]
  \centering
  \includegraphics[width=.48\textwidth]{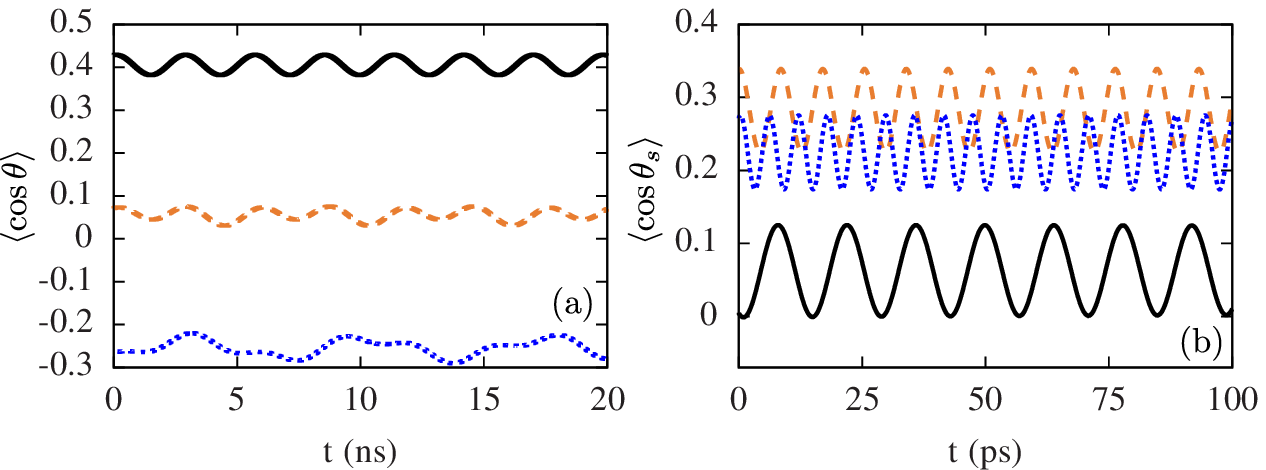}
  \caption{(Color online) 
Orientation cosines  once the peak intensity and dc field strength are kept constant for $t>0$.
For the state \pstate{1}{1}{e}, 
(a) expectation value $\langle\cos \theta\rangle$ with $\Ialign=\SI{2e11}{\intensity}$ and 
$\beta=0\degree$ (solid), 
$30\degree$ (dashed) and
$45\degree$ (dotted);
and (b) $\langle\cos\theta_\textup{s}\rangle$
for $\beta=30\degree$ and 
 $\Ialign=\SI{2e11}{\intensity}$ (solid), 
$\SI{5e11}{\intensity}$ (dashed)
and $\SI{e12}{\intensity}$ (dotted).
The dc field is $E_\textup{s}=\SI{300}{\fieldstrength}$.
}
\label{fig:fig_15}
\end{figure}
Let us first analyze the case when the dynamics takes place within a pendular doublet.
If the adiabatic states are not fully oriented, 
the coupling term in \autoref{eq:exp_value} is non-zero and
this expectation value oscillates  for $t>0$ with the frequency 
equal to the energy splitting of the
corresponding  pendular doublet. 
For the \pstate{1}{1}{e} state,
this behavior is shown for the time evolution of  
$\langle\cos\theta\rangle$ 
in \autoref{fig:fig_15}~(a),
with $\Ialign=\SI{2e11}{\intensity}$, $\textup{E}_\textup{s}=\SI{300}{\fieldstrength}$ 
and $\beta=0\degree$.
An analogous behavior is obtained for the ground state and $0\degree<\beta<90\degree$.
By further increasing the peak intensity, the orientation of the adiabatic states increases, 
the coupling terms are reduced approaching zero, and
these regular oscillations will disappear. 

When two pendular doublets participate in the dynamics,
this oscillatory behavior becomes irregular, because 
the frequencies associated with the energy separations 
within each pendular doublet and between two of them do not form a
commensurable set. 
As an example, we show in
\autoref{fig:fig_15}~(a) these irregular oscillations of $\langle\cos\theta\rangle$
for \pstate{1}{1}{e}  
with $\beta=30\degree$ and $45\degree$, $\textup{E}_\textup{s}=\SI{300}{\fieldstrength}$ and
$\Ialign=\SI{2e11}{\intensity}$.   
By increasing $\Ialign$, the dynamics of this state still has contributions from
different pendular doublets, but the
two states in a pendular pair are not populated. 
As a consequence,  the coupling terms are reduced and the oscillation decreases or even disappears. 

For $0\degree<\beta<90\degree$, the time evolution of $\langle\cos\theta_{\textup{s}}\rangle$
is dominated by the couplings of adiabatic levels from doublets with $|\Delta M|\approx1$.
This is illustrated in  \autoref{fig:fig_15}~(b), for the state \pstate{1}{1}{e},
 $\Ialign=\SI{2e11}{\intensity}$,  $\SI{5e11}{\intensity}$ and  $\SI{e12}{\intensity}$
and  $\beta=30\degree$. 
Independently of $\Ialign$, in this time scale $\langle\cos\theta_{\textup{s}}\rangle$ oscillates 
with the largest frequency given by the energy gap between of the two pendular doublets involved,
which is similar for the three peak intensities.
On a larger time scale, the frequencies due to the
states in a doublet will modulate the oscillations of $\langle\cos\theta_\textup{s}\rangle$
in the weak field regime.


\section{Switching on the laser pulse first: Orientation of the aligned pendular states.}
\label{sec:ordering}
In previous sections, the field configuration was based on the mixed-field orientation 
experiments~\cite{kupper:prl102,kupper:jcp131,nielsen:prl2012}.  
Here, we investigate the molecular dynamics 
when the temporal order of the fields
is inverted: the Gaussian pulse is switched on first, its  peak intensity is kept constant, and
then the static electric field is turned on.  
While the laser field is switched on,  the pendular doublets of quasi-degenerate states 
with opposite parity are formed.
This process is adiabatic~\cite{ortigoso:jcp110,viftrup:prl99}, and for a sufficiently large  peak intensity, 
these two levels are strongly aligned but not oriented. 
By turning on the static field,
these states have the same symmetry and they should be oriented due to their
interaction with this field. 
For this field configuration, 
we check now  the validity of the adiabatic predictions~\cite{friedrich:jcp111,friedrich:jpca103} 
by comparing them to  a time dependent analysis.

The peak intensity of the Gaussian pulse 
is reached at $t=0$ and kept constant afterwards.
At this point, if $\Ialign$ is large enough, the energy gap between the states in a pendular doublet is much smaller than 
the energy gap with the neighboring doublet. 
Then, for a certain pendular level, its rotational dynamics for $t>0$, \ie,
 when the static field is switched on, could 
be approximated by a two-state model involving  the two levels forming the corresponding pendular 
doublet~\cite{friedrich:jpca103}.
At $t=0$, \ie,  $\Estatabs(0)=0$ and $\textup{I}(0)=\Ialign$, 
the pendular states are $|\psi_{l}\rangle$ with $l=e$ and $o$ indicating even or odd parity. 
Under this approximation, we assume that the levels  $|\psi_r\rangle=\frac{1}{\sqrt{2}}\left(|\psi_e\rangle+|\psi_o\rangle\right)$
and
$|\psi_w\rangle=\frac{1}{\sqrt{2}}\left(|\psi_e\rangle-|\psi_o\rangle\right)$ are 
right- and wrong-way oriented, respectively. 
The two-state-model Hamiltonian yields as
\begin{equation*}
  \label{eq:two_state_ham}
H(t)=\left(  \begin{array}{cc}
  0 & -\mu v_{\textup{s}}t  \langle\cos\theta_{\textup{s}}\rangle_{eo}\\
-\mu v_{\textup{s}}t  \langle\cos\theta_{\textup{s}}\rangle_{eo} & \Delta E
  \end{array}\right),
\end{equation*}
where we have taken 
$\Estatabs(t)=v_\textup{s} t$
with $v_\textup{s}=\Estatabs/T_0$ and 
$T_0$ being the switching on speed and time, respectively. 
This time $T_0$ is chosen so that if these states are exposed only to this  field, the 
turning-on process is  adiabatic. Note that we have taken 
 $\langle\psi_e|H|\psi_e\rangle=0$,  $\langle\psi_o|H|\psi_o\rangle=\Delta E$,
 and 
$\langle\psi_{e}|H|\psi_o\rangle=\langle\psi_{o}|H|\psi_e\rangle
=-\mu v_{\textup{s}}t\langle\psi_e|\cos\theta_{\textup{s}}|\psi_o\rangle=-\mu v_{\textup{s}}t\langle\cos\theta_{\textup{s}}\rangle_{eo}$.
The time-dependent
Schr\"odinger equation associated to this Hamiltonian  
admits a scaling factor. That is, 
when the dynamics is adiabatic using $v_\textup{s}$
for a pendular doublet with energy splitting 
 $\Delta E$ at $\textup{I}(0)=\Ialign$,  
then, for a peak intensity $\Ialign'$ 
and  $\Delta E^\prime=k \Delta E$, 
the dynamics  is adiabatic for  $v_\textup{s}^\prime=k^2 v_\textup{s}$.

\begin{figure}[th]
    \centering
\includegraphics[width=.5\textwidth]{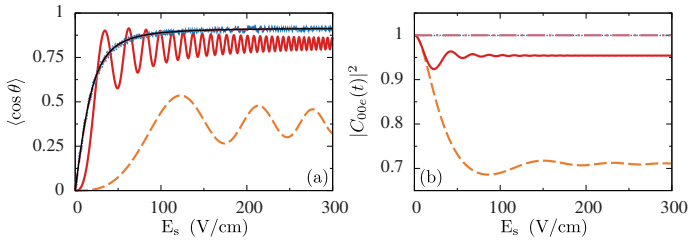}
 \caption{(Color online) For the ground state,  (a) expectation value  $\langle\cos\theta\rangle$ 
and (b) weight of the adiabatic ground state on 
its time-dependent wave function as 
a function of  $\Estatabs(t)$, for turning on speeds 
 $v_\textup{s}=10^{10}$ (orange dashed),  $10^9$ (red solid), 
 $10^8$ (blue dotted) and $10^7$~V/scm (pink dot-dashed), and adiabatic
results (thin solid)
 The fields are parallel and  the Gaussian pulse has $\tau=10$~ns and 
  $\Ialign=\SI{2e11}{\intensity}$.}
 \label{fig:fig_16}
 \end{figure}
For the sake of simplicity, we focus 
on the ground state in a parallel-field configuration.
For several switching on
speeds,  \autoref{fig:fig_16}~(a) and (b)
display the directional cosine and 
the population of the adiabatic ground state,  respectively, as
a function of $\Estatabs(t)$. 
The Gaussian pulse has $10$~ns FWHM and peak intensity  
$\Ialign=\SI{2e11}{\intensity}$. 
Before the dc field is turned on,  
the alignment of the ground state is $\langle\cos^2\theta\rangle=0.845$,
the energy separation within this pendular pair is
$\Delta E\approx 5.36\times 10^{-4}$~cm$^{-1}$,
and there are $2.4$~cm$^{-1}$ to the next  pendular doublet.
For $\Estatabs=\SI{1}{\fieldstrength}$, 
the coupling term is 
$\mu\langle\cos\theta\rangle_{eo}\Estatabs=1.09\times 10^{-5}$~cm$^{-1}$ 
with $\langle\cos\theta\rangle_{eo}=0.915$.
In an adiabatic picture, 
the energy gap $\Delta E$
can not be neglected, and 
as $\Estatabs(t)$ 
is increased the energy of the ground state does not increase linearly 
with $\Estatabs$~\cite{friedrich:jpca103}. 
For $v_\textup{s}=10^{10}$~V/scm, the adiabatic parameter $\eta\approx 1$, 
the rotational dynamics is non-adiabatic and there is a population transfer between
the two states in this doublet. 
We note that for this process, the adiabatic parameter $\eta$ is defined as 
in \autoref{eq:adiabatic_criteria} but replacing the laser field 
interaction $H_\textup{L}(t)$ \eqref{eq:hl} by the dipole term $H_{\textup{s}}(t)$ \eqref{eq:hs}. 
The ground state presents a moderate orientation bellow the adiabatic limit
due to the contributions of the  adiabatic states \ppstate{0}{0}{e} and \ppstate{1}{0}{e},
 $|C_{00e}(t)|^2$ decreases until a minimum value  showing a smooth oscillation afterwards, cf.  \autoref{fig:fig_16}~(b). 
Due to the coupling term,  $\langle\cos\theta\rangle$ oscillates as $\Estatabs(t)$ is increased,  
and its frequency is equal to
the energy separation between the adiabatic levels  \ppstate{0}{0}{e} and \ppstate{1}{0}{e}. 
A similar behavior is observed for  $10^{9}$~V/scm, but the orientation of the ground state 
oscillates around a value closer
to the adiabatic limit because the process  is more adiabatic and 
$|C_{00e}(t)|^2\approx0.956$  for $\Estatabs(t) \gtrsim\SI{100}{\fieldstrength}$.   
For $v_\textup{s}=10^{8}$ and $10^{7}$~V/scm, the dynamics can be
considered as adiabatic, with $|C_{00e}(t)|^2$ being larger than $0.999$. 
However, for $v_\textup{s}=10^{8}$,  $\langle\cos\theta\rangle$ still oscillates
around the adiabatic value.

By increasing the peak intensity of the laser pulse, the energy  splitting 
of the  levels in a pendular doublet
is decreased, but their coupling due to the dc field is not significantly modified. Thus, 
the rotational dynamics becomes more diabatic, and larger turning-on times are needed to 
achieve the adiabatic limit. 
For $\Ialign=\SI{5e11}{\intensity}$,
the ground state is separated by $\Delta E\approx7.7\times 10^{-7}$~cm$^{-1}$ from \ppstate{1}{0}{e},  and 
by $3.9$~cm$^{-1}$ from  the next pendular doublet. The coupling 
due to the dc field is
$\mu\langle\cos\theta\rangle_{eo}\Estatabs=1.13\times 10^{-5}$~cm$^{-1}$ 
for $\Estatabs=\SI{1}{\fieldstrength}$ and with
$\langle\cos\theta\rangle_{eo}=0.948$.  According to the scaling law of the time-dependent 
Schr\"odinger equation, this process would be
adiabatic for a speed of $v_\textup{s}\lesssim 20$~V/scm.

For $\Ialign=\SI{e12}{\intensity}$ and the ground state, 
we find $\Delta E\approx 3.3\times 10^{-10}$~cm$^{-1}$, 
$5.7$~cm$^{-1}$ to the second doublet, and
$\mu\langle\cos\theta\rangle_{eo}\Estatabs=1.18\times 10^{-5}$~cm$^{-1}$ 
for $\Estatabs=\SI{1}{\fieldstrength}$ and with
$\langle\cos\theta\rangle_{eo}=0.964$. 
Within an adiabatic framework, as $\Estatabs$ is increased 
the  ground state  energy 
can be approximated by the pseudo-first-order Stark linear effect 
$|\mu \Estatabs\, $\rppstate{0}{0}{e}$\cos\theta$\ppstate{1}{0}{e}$|$~\cite{friedrich:jpca103}.
Note that $\Delta E$ is smaller than the dc filed coupling even for 
 $\Estatabs\approx \SI{e-4}{\fieldstrength}$.
Based on the scaling law of the two-state model Schr\"odinger equation, 
the dc field should be turned on very slowly,  $v_\textup{s}\lesssim 10^{-2}$~V/scm,
to achieve the adiabatic limit.
For larger turning-on speeds, 
the dynamics is so diabatic that the \pstate{0}{0}{e} wave function 
does not change, and its projections on the  adiabatic states  \ppstate{0}{0}{e} and \ppstate{1}{0}{e}
are close to the field-free values even for $v_\textup{s}\approx 10^5$~V/scm. 

\section{Conclusions}
\label{sec:conclusions}


In this work, we have investigated the mixed-field orientation dynamics of linear molecules. 
The richness and variety of the field-dressed rotational dynamics has been
illustrated by analyzing in detail the directional properties of several low-lying states. 
In particular, we have explored the degree of orientation as the 
peak intensity and FWHM of the Gaussian pulse, the electrostatic field strength and the angle 
between both fields are varied.

By considering  prototypical field configurations used in current mixed-field orientation   experiments, 
we have proven that the assumption of a fully adiabatic dynamics is incorrect.
For parallel fields, a non-adiabatic transfer of population takes
place when the quasi-degenerated pendular doublets are formed 
as the laser intensity is increased. 
As a consequence, the time-dependent results for the  degree of orientation are smaller
than the predictions of the adiabatic theory. Using current available experimental peak intensities,
longer laser pulses or stronger static fields will increase the degree of orientation even
for highly excited  states. In particular, we have provided the field parameters under which
the mixed-field orientation dynamics will be fully adiabatic.
We have also shown that the field-dressed dynamics is more complicated if both fields are tilted. 
Apart from the non-adiabatic effects when the pendular doublets are formed, at weak laser
intensities there is also population transfer due to the splitting of the states within a $J$-manifold having now
the same symmetry. For non-parallel fields, we have shown that the ground state is strongly oriented,
whereas excited states might only present a moderate orientation, and, furthermore, some of them
could behave as dark states to the mixed-field orientation dynamics.
The requirements for an adiabatic dynamics are now more difficult to satisfy for excited levels
than for the ground state. Again, we have indicated the field configuration that will give rise to
an adiabatic mixed-field-orientation. 
If the peak intensity  is kept constant after turning on the pulse, we have shown that 
the orientation of the states might oscillate with time due to the non-adiabatic dynamics.
Finally, we have  investigated the molecular dynamics when the temporal order of the fields is
inverted. We have shown that once the ground state is  adiabatically aligned,
the switching on of the dc field has to be very slow to achieve a significant orientation.

Although our study is restricted to the OCS molecule, we stress that the above-observed physical
phenomena are expected to occur in many other polar
molecules. Indeed, the Hamiltonian can be
  rescaled, and the above results used to describe another polar linear molecule.
In addition, due the complexity of their rotational level structure of asymmetric tops, 
these non-adiabatic effects should have a negative impact in their
mixed-field orientation experiments~\cite{kupper:prl102,omiste:pccp2011}.

\begin{acknowledgments}
We would like to thank B. Friedrich, J. K\"upper, J. H. Nielsen, P. Schmelcher and H. Stapelfeldt  for
fruitful discussions.  
Financial support by the Spanish project FIS2011-24540 (MICINN), the
Grants FQM-2445 and FQM-4643 (Junta de Andaluc\'{\i}a), and Andalusian research group FQM-207 is
gratefully appreciated. J.J.O. acknowledges the support of ME under the program FPU.

\end{acknowledgments}

\bibliographystyle{apsrev}

\begin{thebibliography}{42}
\expandafter\ifx\csname natexlab\endcsname\relax\def\natexlab#1{#1}\fi
\expandafter\ifx\csname bibnamefont\endcsname\relax
  \def\bibnamefont#1{#1}\fi
\expandafter\ifx\csname bibfnamefont\endcsname\relax
  \def\bibfnamefont#1{#1}\fi
\expandafter\ifx\csname citenamefont\endcsname\relax
  \def\citenamefont#1{#1}\fi
\expandafter\ifx\csname url\endcsname\relax
  \def\url#1{\texttt{#1}}\fi
\expandafter\ifx\csname urlprefix\endcsname\relax\def\urlprefix{URL }\fi
\providecommand{\bibinfo}[2]{#2}
\providecommand{\eprint}[2][]{\url{#2}}

\bibitem[{\citenamefont{Brooks}(1976)}]{brooks:science}
\bibinfo{author}{\bibfnamefont{P.~R.} \bibnamefont{Brooks}},
  \bibinfo{journal}{Science} \textbf{\bibinfo{volume}{193}},
  \bibinfo{pages}{11} (\bibinfo{year}{1976}).

\bibitem[{\citenamefont{Brooks and Jones}(1966)}]{brooks:jcp45}
\bibinfo{author}{\bibfnamefont{P.~R.} \bibnamefont{Brooks}} \bibnamefont{and}
  \bibinfo{author}{\bibfnamefont{E.~M.} \bibnamefont{Jones}},
  \bibinfo{journal}{J. Chem. Phys.} \textbf{\bibinfo{volume}{45}},
  \bibinfo{pages}{3449} (\bibinfo{year}{1966}).

\bibitem[{\citenamefont{Loesch and M\"oller}(1992)}]{loesch:9016}
\bibinfo{author}{\bibfnamefont{H.~J.} \bibnamefont{Loesch}} \bibnamefont{and}
  \bibinfo{author}{\bibfnamefont{J.}~\bibnamefont{M\"oller}},
  \bibinfo{journal}{J. Chem. Phys.} \textbf{\bibinfo{volume}{97}},
  \bibinfo{pages}{9016} (\bibinfo{year}{1992}).

\bibitem[{\citenamefont{Aoiz et~al.}(1998)\citenamefont{Aoiz, Friedrich,
  Herrero, R\'abanos, and Verdasco}}]{aoiz:chem_phys_lett_289}
\bibinfo{author}{\bibfnamefont{F.~J.} \bibnamefont{Aoiz}},
  \bibinfo{author}{\bibfnamefont{B.}~\bibnamefont{Friedrich}},
  \bibinfo{author}{\bibfnamefont{V.~J.} \bibnamefont{Herrero}},
  \bibinfo{author}{\bibfnamefont{V.~S.} \bibnamefont{R\'abanos}},
  \bibnamefont{and} \bibinfo{author}{\bibfnamefont{J.~E.}
  \bibnamefont{Verdasco}}, \bibinfo{journal}{Chem. Phys. Lett.}
  \textbf{\bibinfo{volume}{289}}, \bibinfo{pages}{132 } (\bibinfo{year}{1998}).

\bibitem[{\citenamefont{Aquilanti et~al.}(2005)\citenamefont{Aquilanti,
  Bartolomei, Pirani, Cappelletti, and Vecchiocattivi}}]{aquilanti:pccp_7}
\bibinfo{author}{\bibfnamefont{V.}~\bibnamefont{Aquilanti}},
  \bibinfo{author}{\bibfnamefont{M.}~\bibnamefont{Bartolomei}},
  \bibinfo{author}{\bibfnamefont{F.}~\bibnamefont{Pirani}},
  \bibinfo{author}{\bibfnamefont{D.}~\bibnamefont{Cappelletti}},
  \bibnamefont{and}
  \bibinfo{author}{\bibfnamefont{F.}~\bibnamefont{Vecchiocattivi}},
  \bibinfo{journal}{Phys. Chem. Chem. Phys} \textbf{\bibinfo{volume}{7}},
  \bibinfo{pages}{291} (\bibinfo{year}{2005}).

\bibitem[{\citenamefont{Bisgaard et~al.}(2009)\citenamefont{Bisgaard, Clarkin,
  Wu, Lee, Gessner, Hayden, and Stolow}}]{Bisgaard:Science323:1464}
\bibinfo{author}{\bibfnamefont{C.~Z.} \bibnamefont{Bisgaard}},
  \bibinfo{author}{\bibfnamefont{O.~J.} \bibnamefont{Clarkin}},
  \bibinfo{author}{\bibfnamefont{G.}~\bibnamefont{Wu}},
  \bibinfo{author}{\bibfnamefont{A.~M.~D.} \bibnamefont{Lee}},
  \bibinfo{author}{\bibfnamefont{O.}~\bibnamefont{Gessner}},
  \bibinfo{author}{\bibfnamefont{C.~C.} \bibnamefont{Hayden}},
  \bibnamefont{and} \bibinfo{author}{\bibfnamefont{A.}~\bibnamefont{Stolow}},
  \bibinfo{journal}{Science} \textbf{\bibinfo{volume}{323}},
  \bibinfo{pages}{1464} (\bibinfo{year}{2009}).

\bibitem[{\citenamefont{Holmegaard et~al.}(2010)\citenamefont{Holmegaard,
  Hansen, Kalhoj, Kragh, Stapelfeldt, Filsinger, K\"upper, Meijer, Dimitrovski,
  Abu-samha et~al.}}]{Holmegaard:natphys6}
\bibinfo{author}{\bibfnamefont{L.}~\bibnamefont{Holmegaard}},
  \bibinfo{author}{\bibfnamefont{J.~L.} \bibnamefont{Hansen}},
  \bibinfo{author}{\bibfnamefont{L.}~\bibnamefont{Kalhoj}},
  \bibinfo{author}{\bibfnamefont{S.~L.} \bibnamefont{Kragh}},
  \bibinfo{author}{\bibfnamefont{H.}~\bibnamefont{Stapelfeldt}},
  \bibinfo{author}{\bibfnamefont{F.}~\bibnamefont{Filsinger}},
  \bibinfo{author}{\bibfnamefont{J.}~\bibnamefont{K\"upper}},
  \bibinfo{author}{\bibfnamefont{G.}~\bibnamefont{Meijer}},
  \bibinfo{author}{\bibfnamefont{D.}~\bibnamefont{Dimitrovski}},
  \bibinfo{author}{\bibfnamefont{M.}~\bibnamefont{Abu-samha}},
  \bibnamefont{et~al.}, \bibinfo{journal}{Nat. Phys.}
  \textbf{\bibinfo{volume}{6}}, \bibinfo{pages}{428} (\bibinfo{year}{2010}).

\bibitem[{\citenamefont{Hansen et~al.}(2011)\citenamefont{Hansen, Stapelfeldt,
  Dimitrovski, Abu-samha, Martiny, and Madsen}}]{Hansen:PRL106:073001}
\bibinfo{author}{\bibfnamefont{J.~L.} \bibnamefont{Hansen}},
  \bibinfo{author}{\bibfnamefont{H.}~\bibnamefont{Stapelfeldt}},
  \bibinfo{author}{\bibfnamefont{D.}~\bibnamefont{Dimitrovski}},
  \bibinfo{author}{\bibfnamefont{M.}~\bibnamefont{Abu-samha}},
  \bibinfo{author}{\bibfnamefont{C.~P.~J.} \bibnamefont{Martiny}},
  \bibnamefont{and} \bibinfo{author}{\bibfnamefont{L.~B.}
  \bibnamefont{Madsen}}, \bibinfo{journal}{Phys. Rev. Lett.}
  \textbf{\bibinfo{volume}{106}}, \bibinfo{pages}{073001}
  (\bibinfo{year}{2011}).

\bibitem[{\citenamefont{Frumker et~al.}()\citenamefont{Frumker, Hebeisen,
  Kajumba, J.~B. Bertrand~and, Spanner, Villeneuve, Naumov, and
  Corkum}}]{frumker2012}
\bibinfo{author}{\bibfnamefont{E.}~\bibnamefont{Frumker}},
  \bibinfo{author}{\bibfnamefont{C.~T.} \bibnamefont{Hebeisen}},
  \bibinfo{author}{\bibfnamefont{N.}~\bibnamefont{Kajumba}},
  \bibinfo{author}{\bibfnamefont{H.~J.~W.} \bibnamefont{J.~B. Bertrand~and}},
  \bibinfo{author}{\bibfnamefont{M.}~\bibnamefont{Spanner}},
  \bibinfo{author}{\bibfnamefont{D.~M.} \bibnamefont{Villeneuve}},
  \bibinfo{author}{\bibfnamefont{A.}~\bibnamefont{Naumov}}, \bibnamefont{and}
  \bibinfo{author}{\bibfnamefont{P.}~\bibnamefont{Corkum}},
  \bibinfo{note}{arXiv:1205.4199v1}.

\bibitem[{\citenamefont{Kraus et~al.}(2012)\citenamefont{Kraus, Vlajkovic,
  Rupenyan, and W\"orner}}]{kraus2012}
\bibinfo{author}{\bibfnamefont{P.~M.} \bibnamefont{Kraus}},
  \bibinfo{author}{\bibfnamefont{S.}~\bibnamefont{Vlajkovic}},
  \bibinfo{author}{\bibfnamefont{A.}~\bibnamefont{Rupenyan}}, \bibnamefont{and}
  \bibinfo{author}{\bibfnamefont{H.~J.} \bibnamefont{W\"orner}},
  \bibinfo{journal}{Phys. Rev. Lett.} \textbf{\bibinfo{volume}{accepted}}
  (\bibinfo{year}{2012}).

\bibitem[{\citenamefont{Stolte}(1988)}]{stolte}
\bibinfo{author}{\bibfnamefont{S.}~\bibnamefont{Stolte}},
  \emph{\bibinfo{title}{Atomic and Molecular Beam Methods}}
  (\bibinfo{publisher}{Oxford University Press, New York},
  \bibinfo{year}{1988}).

\bibitem[{\citenamefont{Loesch and Remscheid}(1990)}]{loesch:jcp93}
\bibinfo{author}{\bibfnamefont{H.~J.} \bibnamefont{Loesch}} \bibnamefont{and}
  \bibinfo{author}{\bibfnamefont{A.}~\bibnamefont{Remscheid}},
  \bibinfo{journal}{J. Chem. Phys.} \textbf{\bibinfo{volume}{93}},
  \bibinfo{pages}{4779} (\bibinfo{year}{1990}).

\bibitem[{\citenamefont{Friedrich and Herschbach}(1991)}]{friedrich:nature353}
\bibinfo{author}{\bibfnamefont{B.}~\bibnamefont{Friedrich}} \bibnamefont{and}
  \bibinfo{author}{\bibfnamefont{D.~R.} \bibnamefont{Herschbach}},
  \bibinfo{journal}{Nature} \textbf{\bibinfo{volume}{353}},
  \bibinfo{pages}{412} (\bibinfo{year}{1991}).

\bibitem[{\citenamefont{Friedrich et~al.}(1991)\citenamefont{Friedrich,
  Pullman, and Herschbach}}]{friedrich:jpc95}
\bibinfo{author}{\bibfnamefont{B.}~\bibnamefont{Friedrich}},
  \bibinfo{author}{\bibfnamefont{D.~P.} \bibnamefont{Pullman}},
  \bibnamefont{and} \bibinfo{author}{\bibfnamefont{D.~R.}
  \bibnamefont{Herschbach}}, \bibinfo{journal}{J. Phys. Chem.}
  \textbf{\bibinfo{volume}{95}}, \bibinfo{pages}{8118} (\bibinfo{year}{1991}).

\bibitem[{\citenamefont{Block et~al.}(1992)\citenamefont{Block, Bohac, and
  Miller}}]{block:prl68}
\bibinfo{author}{\bibfnamefont{P.~A.} \bibnamefont{Block}},
  \bibinfo{author}{\bibfnamefont{E.~J.} \bibnamefont{Bohac}}, \bibnamefont{and}
  \bibinfo{author}{\bibfnamefont{R.~E.} \bibnamefont{Miller}},
  \bibinfo{journal}{Phys. Rev. Lett.} \textbf{\bibinfo{volume}{68}},
  \bibinfo{pages}{1303} (\bibinfo{year}{1992}).

\bibitem[{\citenamefont{Friedrich et~al.}(1992)\citenamefont{Friedrich, Rubahn,
  and Sathyamurthy}}]{friedrich:prl69}
\bibinfo{author}{\bibfnamefont{B.}~\bibnamefont{Friedrich}},
  \bibinfo{author}{\bibfnamefont{H.-G.} \bibnamefont{Rubahn}},
  \bibnamefont{and}
  \bibinfo{author}{\bibfnamefont{N.}~\bibnamefont{Sathyamurthy}},
  \bibinfo{journal}{Phys. Rev. Lett.} \textbf{\bibinfo{volume}{69}},
  \bibinfo{pages}{2487} (\bibinfo{year}{1992}).

\bibitem[{\citenamefont{Slenczka et~al.}(1994)\citenamefont{Slenczka,
  Friedrich, and Herschbach}}]{slenczka:prl72}
\bibinfo{author}{\bibfnamefont{A.}~\bibnamefont{Slenczka}},
  \bibinfo{author}{\bibfnamefont{B.}~\bibnamefont{Friedrich}},
  \bibnamefont{and}
  \bibinfo{author}{\bibfnamefont{D.}~\bibnamefont{Herschbach}},
  \bibinfo{journal}{Phys. Rev. Lett.} \textbf{\bibinfo{volume}{72}},
  \bibinfo{pages}{1806} (\bibinfo{year}{1994}).

\bibitem[{\citenamefont{Friedrich and
  Herschbach}(1999{\natexlab{a}})}]{friedrich:jcp111}
\bibinfo{author}{\bibfnamefont{B.}~\bibnamefont{Friedrich}} \bibnamefont{and}
  \bibinfo{author}{\bibfnamefont{D.~R.} \bibnamefont{Herschbach}},
  \bibinfo{journal}{J. Chem. Phys.} \textbf{\bibinfo{volume}{111}},
  \bibinfo{pages}{6157} (\bibinfo{year}{1999}{\natexlab{a}}).

\bibitem[{\citenamefont{Friedrich and
  Herschbach}(1999{\natexlab{b}})}]{friedrich:jpca103}
\bibinfo{author}{\bibfnamefont{B.}~\bibnamefont{Friedrich}} \bibnamefont{and}
  \bibinfo{author}{\bibfnamefont{D.}~\bibnamefont{Herschbach}},
  \bibinfo{journal}{J. Phys. Chem. A} \textbf{\bibinfo{volume}{103}},
  \bibinfo{pages}{10280} (\bibinfo{year}{1999}{\natexlab{b}}).

\bibitem[{\citenamefont{Sakai et~al.}(2003)\citenamefont{Sakai, Minemoto,
  Nanjo, Tanji, and Suzuki}}]{sakai:prl_90}
\bibinfo{author}{\bibfnamefont{H.}~\bibnamefont{Sakai}},
  \bibinfo{author}{\bibfnamefont{S.}~\bibnamefont{Minemoto}},
  \bibinfo{author}{\bibfnamefont{H.}~\bibnamefont{Nanjo}},
  \bibinfo{author}{\bibfnamefont{H.}~\bibnamefont{Tanji}}, \bibnamefont{and}
  \bibinfo{author}{\bibfnamefont{T.}~\bibnamefont{Suzuki}},
  \bibinfo{journal}{Phys. Rev. Lett.} \textbf{\bibinfo{volume}{90}},
  \bibinfo{pages}{083001} (\bibinfo{year}{2003}).

\bibitem[{\citenamefont{Buck and F\'arn\'ik}(2006)}]{Buck:IRPC25:583}
\bibinfo{author}{\bibfnamefont{U.}~\bibnamefont{Buck}} \bibnamefont{and}
  \bibinfo{author}{\bibfnamefont{M.}~\bibnamefont{F\'arn\'ik}},
  \bibinfo{journal}{Int.\ Rev.\ Phys.\ Chem.} \textbf{\bibinfo{volume}{25}},
  \bibinfo{pages}{583} (\bibinfo{year}{2006}).

\bibitem[{\citenamefont{Holmegaard et~al.}(2009)\citenamefont{Holmegaard,
  Nielsen, Nevo, Stapelfeldt, Filsinger, K\"upper, and Meijer}}]{kupper:prl102}
\bibinfo{author}{\bibfnamefont{L.}~\bibnamefont{Holmegaard}},
  \bibinfo{author}{\bibfnamefont{J.~H.} \bibnamefont{Nielsen}},
  \bibinfo{author}{\bibfnamefont{I.}~\bibnamefont{Nevo}},
  \bibinfo{author}{\bibfnamefont{H.}~\bibnamefont{Stapelfeldt}},
  \bibinfo{author}{\bibfnamefont{F.}~\bibnamefont{Filsinger}},
  \bibinfo{author}{\bibfnamefont{J.}~\bibnamefont{K\"upper}}, \bibnamefont{and}
  \bibinfo{author}{\bibfnamefont{G.}~\bibnamefont{Meijer}},
  \bibinfo{journal}{Phys. Rev. Lett.} \textbf{\bibinfo{volume}{102}},
  \bibinfo{pages}{023001} (\bibinfo{year}{2009}).

\bibitem[{\citenamefont{Ghafur et~al.}(2009)\citenamefont{Ghafur, Rouzee,
  Gijsbertsen, Siu, Stolte, and Vrakking}}]{ghafur_impulsive_2009}
\bibinfo{author}{\bibfnamefont{O.}~\bibnamefont{Ghafur}},
  \bibinfo{author}{\bibfnamefont{A.}~\bibnamefont{Rouzee}},
  \bibinfo{author}{\bibfnamefont{A.}~\bibnamefont{Gijsbertsen}},
  \bibinfo{author}{\bibfnamefont{W.~K.} \bibnamefont{Siu}},
  \bibinfo{author}{\bibfnamefont{S.}~\bibnamefont{Stolte}}, \bibnamefont{and}
  \bibinfo{author}{\bibfnamefont{M.~J.~J.} \bibnamefont{Vrakking}},
  \bibinfo{journal}{Nat Phys} \textbf{\bibinfo{volume}{5}},
  \bibinfo{pages}{289} (\bibinfo{year}{2009}).

\bibitem[{\citenamefont{Filsinger et~al.}(2009)\citenamefont{Filsinger,
  K\"upper, Meijer, Holmegaard, Nielsen, Nevo, Hansen, and
  Stapelfeldt}}]{kupper:jcp131}
\bibinfo{author}{\bibfnamefont{F.}~\bibnamefont{Filsinger}},
  \bibinfo{author}{\bibfnamefont{J.}~\bibnamefont{K\"upper}},
  \bibinfo{author}{\bibfnamefont{G.}~\bibnamefont{Meijer}},
  \bibinfo{author}{\bibfnamefont{L.}~\bibnamefont{Holmegaard}},
  \bibinfo{author}{\bibfnamefont{J.~H.} \bibnamefont{Nielsen}},
  \bibinfo{author}{\bibfnamefont{I.}~\bibnamefont{Nevo}},
  \bibinfo{author}{\bibfnamefont{J.~L.} \bibnamefont{Hansen}},
  \bibnamefont{and}
  \bibinfo{author}{\bibfnamefont{H.}~\bibnamefont{Stapelfeldt}},
  \bibinfo{journal}{J. Chem. Phys.} \textbf{\bibinfo{volume}{131}},
  \bibinfo{pages}{064309} (\bibinfo{year}{2009}).

\bibitem[{\citenamefont{Omiste et~al.}(2011)\citenamefont{Omiste, G\"arttner,
  Schmelcher, Gonz\'{a}lez-F\'{e}rez, Holmegaard, Nielsen, Stapelfeldt, and
  K\"upper}}]{omiste:pccp2011}
\bibinfo{author}{\bibfnamefont{J.~J.} \bibnamefont{Omiste}},
  \bibinfo{author}{\bibfnamefont{M.}~\bibnamefont{G\"arttner}},
  \bibinfo{author}{\bibfnamefont{P.}~\bibnamefont{Schmelcher}},
  \bibinfo{author}{\bibfnamefont{R.}~\bibnamefont{Gonz\'{a}lez-F\'{e}rez}},
  \bibinfo{author}{\bibfnamefont{L.}~\bibnamefont{Holmegaard}},
  \bibinfo{author}{\bibfnamefont{J.~H.} \bibnamefont{Nielsen}},
  \bibinfo{author}{\bibfnamefont{H.}~\bibnamefont{Stapelfeldt}},
  \bibnamefont{and} \bibinfo{author}{\bibfnamefont{J.}~\bibnamefont{K\"upper}},
  \bibinfo{journal}{Phys. Chem. Chem. Phys.} \textbf{\bibinfo{volume}{13}},
  \bibinfo{pages}{18815} (\bibinfo{year}{2011}).

\bibitem[{\citenamefont{Nielsen et~al.}(2012)\citenamefont{Nielsen,
  Stapelfeldt, K\"upper, Friedrich, Omiste, and
  Gonz\'alez-F\'erez}}]{nielsen:prl2012}
\bibinfo{author}{\bibfnamefont{J.~H.} \bibnamefont{Nielsen}},
  \bibinfo{author}{\bibfnamefont{H.}~\bibnamefont{Stapelfeldt}},
  \bibinfo{author}{\bibfnamefont{J.}~\bibnamefont{K\"upper}},
  \bibinfo{author}{\bibfnamefont{B.}~\bibnamefont{Friedrich}},
  \bibinfo{author}{\bibfnamefont{J.~J.} \bibnamefont{Omiste}},
  \bibnamefont{and}
  \bibinfo{author}{\bibfnamefont{R.}~\bibnamefont{Gonz\'alez-F\'erez}},
  \bibinfo{journal}{Phys. Rev. Lett.} \textbf{\bibinfo{volume}{108}},
  \bibinfo{pages}{193001} (\bibinfo{year}{2012}).

\bibitem[{\citenamefont{Dion et~al.}(1999)\citenamefont{Dion, Keller, Atabek,
  and Bandrauk}}]{dion_pra59}
\bibinfo{author}{\bibfnamefont{C.~M.} \bibnamefont{Dion}},
  \bibinfo{author}{\bibfnamefont{A.}~\bibnamefont{Keller}},
  \bibinfo{author}{\bibfnamefont{O.}~\bibnamefont{Atabek}}, \bibnamefont{and}
  \bibinfo{author}{\bibfnamefont{A.~D.} \bibnamefont{Bandrauk}},
  \bibinfo{journal}{Phys. Rev. A} \textbf{\bibinfo{volume}{59}},
  \bibinfo{pages}{1382} (\bibinfo{year}{1999}).

\bibitem[{\citenamefont{Henriksen}(1999)}]{henriksen_cpl312}
\bibinfo{author}{\bibfnamefont{N.~E.} \bibnamefont{Henriksen}},
  \bibinfo{journal}{Chem. Phys. Lett.} \textbf{\bibinfo{volume}{312}},
  \bibinfo{pages}{196} (\bibinfo{year}{1999}).

\bibitem[{\citenamefont{Feit et~al.}(1982)\citenamefont{Feit, {Fleck Jr.}, and
  Steiger}}]{feit:jcp82}
\bibinfo{author}{\bibfnamefont{M.~D.} \bibnamefont{Feit}},
  \bibinfo{author}{\bibfnamefont{J.~A.} \bibnamefont{{Fleck Jr.}}},
  \bibnamefont{and} \bibinfo{author}{\bibfnamefont{A.}~\bibnamefont{Steiger}},
  \bibinfo{journal}{J. Comp. Phys.} \textbf{\bibinfo{volume}{47}},
  \bibinfo{pages}{412} (\bibinfo{year}{1982}).

\bibitem[{\citenamefont{Ba{\u{c}i\`c} and Light}(1989)}]{bacic:arpc89}
\bibinfo{author}{\bibfnamefont{Z.}~\bibnamefont{Ba{\u{c}i\`c}}}
  \bibnamefont{and} \bibinfo{author}{\bibfnamefont{J.~C.} \bibnamefont{Light}},
  \bibinfo{journal}{Annu. Rev. Phys. Chem.} \textbf{\bibinfo{volume}{40}},
  \bibinfo{pages}{469} (\bibinfo{year}{1989}).

\bibitem[{\citenamefont{Corey and Lemoine}(1992)}]{corey:jcp92}
\bibinfo{author}{\bibfnamefont{G.~C.} \bibnamefont{Corey}} \bibnamefont{and}
  \bibinfo{author}{\bibfnamefont{D.}~\bibnamefont{Lemoine}},
  \bibinfo{journal}{J. Chem. Phys.} \textbf{\bibinfo{volume}{97}},
  \bibinfo{pages}{4115} (\bibinfo{year}{1992}).

\bibitem[{\citenamefont{Offer and Balint-Kurti}(1994)}]{offer:10416}
\bibinfo{author}{\bibfnamefont{A.~R.} \bibnamefont{Offer}} \bibnamefont{and}
  \bibinfo{author}{\bibfnamefont{G.~G.} \bibnamefont{Balint-Kurti}},
  \bibinfo{journal}{J. Chem. Phys.} \textbf{\bibinfo{volume}{101}},
  \bibinfo{pages}{10416} (\bibinfo{year}{1994}).

\bibitem[{\citenamefont{S\'anchez-Moreno
  et~al.}(2007)\citenamefont{S\'anchez-Moreno, Gonz\'alez-F\'erez, and
  Schmelcher}}]{sanchezmoreno:PRA.2007}
\bibinfo{author}{\bibfnamefont{P.}~\bibnamefont{S\'anchez-Moreno}},
  \bibinfo{author}{\bibfnamefont{R.}~\bibnamefont{Gonz\'alez-F\'erez}},
  \bibnamefont{and}
  \bibinfo{author}{\bibfnamefont{P.}~\bibnamefont{Schmelcher}},
  \bibinfo{journal}{Phys. Rev. A} \textbf{\bibinfo{volume}{76}},
  \bibinfo{pages}{053413} (\bibinfo{year}{2007}).

\bibitem[{\citenamefont{H\"artelt and Friedrich}(2008)}]{hartelt_jcp128}
\bibinfo{author}{\bibfnamefont{M.}~\bibnamefont{H\"artelt}} \bibnamefont{and}
  \bibinfo{author}{\bibfnamefont{B.}~\bibnamefont{Friedrich}},
  \bibinfo{journal}{J. Chem. Phys.} \textbf{\bibinfo{volume}{128}},
  \bibinfo{pages}{224313} (\bibinfo{year}{2008}).

\bibitem[{\citenamefont{Ballentine}(1998)}]{ballentine:quantum_mechanics}
\bibinfo{author}{\bibfnamefont{L.~B.} \bibnamefont{Ballentine}},
  \emph{\bibinfo{title}{Quantum {M}echanics: {A} {M}odern {D}evelopment}}
  (\bibinfo{publisher}{World Scientific, Singapore}, \bibinfo{year}{1998}).

\bibitem[{\citenamefont{Friedrich et~al.}(2003)\citenamefont{Friedrich, Nahler,
  and Buck}}]{friedrich:jmodopt50}
\bibinfo{author}{\bibfnamefont{B.}~\bibnamefont{Friedrich}},
  \bibinfo{author}{\bibfnamefont{N.}~\bibnamefont{Nahler}}, \bibnamefont{and}
  \bibinfo{author}{\bibfnamefont{U.}~\bibnamefont{Buck}}, \bibinfo{journal}{J.
  Mod. Opt.} \textbf{\bibinfo{volume}{50}}, \bibinfo{pages}{2677}
  (\bibinfo{year}{2003}).

\bibitem[{\citenamefont{Poulsen et~al.}(2006)\citenamefont{Poulsen, Ejdrup,
  Stapelfeldt, Hamilton, and Seideman}}]{poulsen:phys_rev_a_73}
\bibinfo{author}{\bibfnamefont{M.~D.} \bibnamefont{Poulsen}},
  \bibinfo{author}{\bibfnamefont{T.}~\bibnamefont{Ejdrup}},
  \bibinfo{author}{\bibfnamefont{H.}~\bibnamefont{Stapelfeldt}},
  \bibinfo{author}{\bibfnamefont{E.}~\bibnamefont{Hamilton}}, \bibnamefont{and}
  \bibinfo{author}{\bibfnamefont{T.}~\bibnamefont{Seideman}},
  \bibinfo{journal}{Phys. Rev. A} \textbf{\bibinfo{volume}{73}},
  \bibinfo{pages}{033405} (\bibinfo{year}{2006}).

\bibitem[{\citenamefont{Sugawara et~al.}(2008)\citenamefont{Sugawara, Goban,
  Minemoto, and Sakai}}]{Sugawara2008}
\bibinfo{author}{\bibfnamefont{Y.}~\bibnamefont{Sugawara}},
  \bibinfo{author}{\bibfnamefont{A.}~\bibnamefont{Goban}},
  \bibinfo{author}{\bibfnamefont{S.}~\bibnamefont{Minemoto}}, \bibnamefont{and}
  \bibinfo{author}{\bibfnamefont{H.}~\bibnamefont{Sakai}},
  \bibinfo{journal}{Phys. Rev. A} \textbf{\bibinfo{volume}{77}},
  \bibinfo{pages}{031403(R)} (\bibinfo{year}{2008}).

\bibitem[{\citenamefont{Muramatsu et~al.}(2009)\citenamefont{Muramatsu, Hita,
  Minemoto, and Sakai}}]{Muramatsu2009}
\bibinfo{author}{\bibfnamefont{M.}~\bibnamefont{Muramatsu}},
  \bibinfo{author}{\bibfnamefont{M.}~\bibnamefont{Hita}},
  \bibinfo{author}{\bibfnamefont{S.}~\bibnamefont{Minemoto}}, \bibnamefont{and}
  \bibinfo{author}{\bibfnamefont{H.}~\bibnamefont{Sakai}},
  \bibinfo{journal}{Phys. Rev. A} \textbf{\bibinfo{volume}{79}},
  \bibinfo{pages}{011403(R)} (\bibinfo{year}{2009}).

\bibitem[{\citenamefont{Zare}(1988)}]{zare}
\bibinfo{author}{\bibfnamefont{R.~N.} \bibnamefont{Zare}},
  \emph{\bibinfo{title}{Angular {M}omentum: {U}nderstanding {S}patial {A}spects
  in {C}hemistry and {P}hysics}} (\bibinfo{publisher}{John Wiley and Sons, New
  York}, \bibinfo{year}{1988}).

\bibitem[{\citenamefont{Ortigoso et~al.}(1999)\citenamefont{Ortigoso,
  Rodr\'iguez, Gupta, and Friedrich}}]{ortigoso:jcp110}
\bibinfo{author}{\bibfnamefont{J.}~\bibnamefont{Ortigoso}},
  \bibinfo{author}{\bibfnamefont{M.}~\bibnamefont{Rodr\'iguez}},
  \bibinfo{author}{\bibfnamefont{M.}~\bibnamefont{Gupta}}, \bibnamefont{and}
  \bibinfo{author}{\bibfnamefont{B.}~\bibnamefont{Friedrich}},
  \bibinfo{journal}{J. Chem. Phys.} \textbf{\bibinfo{volume}{110}},
  \bibinfo{pages}{3870} (\bibinfo{year}{1999}).

\bibitem[{\citenamefont{Viftrup et~al.}(2007)\citenamefont{Viftrup, Kumarappan,
  Trippel, Stapelfeldt, Hamilton, and Seideman}}]{viftrup:prl99}
\bibinfo{author}{\bibfnamefont{S.~S.} \bibnamefont{Viftrup}},
  \bibinfo{author}{\bibfnamefont{V.}~\bibnamefont{Kumarappan}},
  \bibinfo{author}{\bibfnamefont{S.}~\bibnamefont{Trippel}},
  \bibinfo{author}{\bibfnamefont{H.}~\bibnamefont{Stapelfeldt}},
  \bibinfo{author}{\bibfnamefont{E.}~\bibnamefont{Hamilton}}, \bibnamefont{and}
  \bibinfo{author}{\bibfnamefont{T.}~\bibnamefont{Seideman}},
  \bibinfo{journal}{Phys. Rev. Lett.} \textbf{\bibinfo{volume}{99}},
  \bibinfo{pages}{143602} (\bibinfo{year}{2007}).

\end{thebibliography}

\end{document}